\documentclass[12pt,preprint]{aastex}
\usepackage{natbib}

\shorttitle{Extinction Corrected SFRs Empirically Derived from UV$-$Optical Colors}
\shortauthors{Treyer et al.}

\begin{document}

\title{Extinction Corrected Star Formation Rates Empirically Derived from Ultraviolet$-$Optical Colors}

\author{
Marie Treyer\altaffilmark{1,2}, 
David Schiminovich\altaffilmark{3}, 
Ben Johnson\altaffilmark{3},
Mark Seibert\altaffilmark{4}, 
Ted Wyder\altaffilmark{2}, 
Tom A. Barlow\altaffilmark{2},
Tim Conrow\altaffilmark{2},
Karl Forster\altaffilmark{2},
Peter G. Friedman\altaffilmark{2},
D. Christopher Martin\altaffilmark{2},
Patrick Morrissey\altaffilmark{2},
Susan G. Neff\altaffilmark{9},
Todd Small\altaffilmark{2},
Luciana Bianchi\altaffilmark{6},
Jose Donas\altaffilmark{1},
Timothy M. Heckman\altaffilmark{5},
Young-Wook Lee\altaffilmark{7},
Barry F. Madore\altaffilmark{4},
Bruno Milliard\altaffilmark{1},
R. Michael Rich\altaffilmark{10},
Alex S. Szalay\altaffilmark{5}
Barry Y. Welsh\altaffilmark{8},
Sukyoung K. Yi \altaffilmark{7}
}

\altaffiltext{1}{Laboratoire d'Astrophysique de Marseille, BP 8, Traverse du Siphon, 13376 Marseille Cedex 12, France}
\altaffiltext{2}{California Institute of Technology, MC 405-47, 1200 E. California Boulevard, Pasadena, CA 91125; treyer@srl.caltech.edu}
\altaffiltext{3}{Department of Astronomy, Columbia University, MC 2457, 550 W. 120 St., NY, NY 10027}
\altaffiltext{4}{Observatories of the Carnegie Institution of Washington, 813 Santa Barbara St., Pasadena, CA 91101}
\altaffiltext{5}{Department of Physics and Astronomy, The Johns Hopkins University, Homewood Campus, Baltimore, MD 21218}
\altaffiltext{6}{Center for Astrophysical Sciences, The Johns Hopkins University, 3400 N. Charles St., Baltimore, MD 21218}
\altaffiltext{7}{Center for Space Astrophysics, Yonsei University, Seoul 120-749, Korea}
\altaffiltext{8}{Space Sciences Laboratory, University of California at Berkeley, 601 Campbell Hall, Berkeley, CA 94720}
\altaffiltext{9}{Laboratory for Astronomy and Solar Physics, NASA Goddard Space Flight Center, Greenbelt, MD 20771}
\altaffiltext{10}{Department of Physics and Astronomy, University of California, Los Angeles, CA 90095}

\begin{abstract}
Using a sample of galaxies from the Sloan Digital Sky Survey spectroscopic catalog
with measured star-formation rates (SFRs) and ultraviolet (UV) photometry from the GALEX 
Medium Imaging Survey, we derived empirical linear correlations between the SFR to UV luminosity ratio and the 
UV$-$optical colors of blue sequence galaxies. The relations provide a simple prescription to correct UV data for
dust attenuation that best reconciles the SFRs derived from UV and emission line data. 
The method breaks down for the red sequence population as well as 
for very blue galaxies such as the local ``supercompact'' UV luminous galaxies and the majority of 
high redshift Lyman Break Galaxies which form a low attenuation sequence of their own. 
\end{abstract}

\section{Introduction}

Although directly tracing young massive stars, the ultraviolet (UV; $\lambda\sim 912-3000$ \AA) 
luminosity of a galaxy is not a straightforward measure of its current star-formation rate (SFR),
nor is in fact any other observable related to new born stars. 
In particular, the galaxy's dust content and past star-formation history (SFH) 
have a significant influence on the interpretation of the observed UV flux in 
terms of current star production. A large, sometimes dominant fraction of the 
UV emission may be obscured by dust and reprocessed at far infrared (FIR) wavelengths.
The UV spectral slope is commonly used to estimate this fraction \citep{Meurer99} but
it is strongly affected by the SFH \citep{Kong04}.
Also because UV emitting stars live long enough for successive generations to coexist, 
the SFH over the past few hundred Myrs must be known to translate the dust corrected 
UV flux into a more instantaneous SFR, such as derived from the galaxy's H$\alpha$ emission.
In the case of a constant SFR and in the absence of dust, the UV luminosity to SFR ratio reaches 
a plateau after $\sim 10^8$ yrs (e.g. Kennicutt 1998), but a strong starburst will cause the 
UV luminosity to scale differently with the SFR. Interpreting the UV emission of 
early-type galaxies is also less straightforward due to contamination by older stars \citep{Ree07}.

Dust obscuration and SFH may be estimated with the help of additional data 
(e.g. the Balmer decrement or far-infrared emission for the dust, the Balmer break 
for the SFH) and of theoretical assumptions. 
However such additional information is not always available or in fact acquirable, 
in particular at high redshifts where SFRs are generally derived from UV and/or infrared (IR) 
photometry with rather large uncertainties. New near-IR spectrographs on 8-10m telescopes 
are now making it possible to detect H$\alpha$ emission and/or continuum breaks at $z>2$ 
\citep{vanDokkum05_1NIRspec,Erb06_Ha, Kriek06_3NIRspec,Kriek06_NIRspec}
but the technique is still limited to prominent features. Detecting  H$\alpha$ at  $z\sim 2$ implies 
a SFR greater than a few M$_{\odot}$ yr$^{-1}$ \citep{Erb06_Ha}, which is not representative
of the whole population \citep{Kriek06_NIRspec}. Optical images which pick up the rest-frame UV 
at  $z>2$  remain the easiest data to obtain.

Here we use medium deep UV photometry from the Galaxy Evolution Explorer (GALEX) and the 
wealth of additional data provided by the Sloan Digital Sky Survey (SDSS) to derive simple  
empirical relations between the observed UV luminosity and the SFR of local star-forming galaxies.
SFRs were derived for tens of thousands of SDSS galaxies using their emission lines and 
state-of-the-art models including a consistent treatment of the dust from the UV to the 
far-IR \citep{B04}. 
We assume these SFRs to be the best possible estimates at the present time, given the quality 
of the spectro-photometric data and the tested reliability of the models, and present an
empirical method to recover them from the UV luminosity of galaxies using their UV$-$optical 
colors. We compare our relations with existing methods, in particular attenuation estimates based 
on the slope of the UV continuum which are commonly used, and investigate their limitations both 
locally and at high redshift. 

The data are summarized in Section 2. In Section 3 we review the relation between the UV luminosity 
and the SFR of star-forming galaxies as well as several published methods for estimating their UV 
attenuation. In Section 4 we present simple empirical color relations that best reconcile the UV data 
with the SFR estimates based on emission line measurements, and discuss their limitations.
Our conclusions are presented in Section 5. 
Throughout the paper we assumed a flat $\Lambda$CDM cosmology with $H_0=70~{\rm km~ s^{-1} Mpc^{-1}}$, 
$\Omega_M=0.3$ and $\Omega_{\Lambda}=0.7$, and a Kroupa IMF \citep{KroupaIMF}.

\section{Data and derived physical quantities}

We select galaxies from the Sloan Digital Sky Survey (SDSS) spectroscopic catalog (Data Release 4;
Adelman-McCarthy 2006\nocite{DR4}) with NUV and FUV photometry from the GALEX Medium Imaging Survey 
(Internal Release 1.1; Martin et al. 2005\nocite{Martin05}, Morissey et al. 2005, 2007).
The UV filters have effective wavelengths of 1528 and 2271 \AA\ respectively. The Medium Imaging
Survey (MIS) has a 5 $\sigma$ detection limit of 22.7 (AB magnitude) in both filters for a typical exposure.
This magnitude limit corresponds to a cut in magnitude error of $\sim 0.1$ in the NUV band and $\sim 0.2$ in 
the FUV band \nocite{Bianchi07} (Bianchi et al. 2007, their Fig. 2).
Our primary sample consists of 23400 SDSS galaxies with $r<17.8$, $z>0.005$, measured H$\alpha$ emission,
aperture corrections less than 1.3 dex (defined as the ratio of the total SFR to the SFR estimated 
within the fiber; see below), and GALEX coverage in the NUV band. Adding FUV coverage reduces 
the sample to 17500 galaxies due to occasional failures of the GALEX FUV detector. Galaxies flagged as AGNs 
in the SDSS MPA/JHU DR4 value-added catalogs\footnote{http://www.mpa-garching.mpg.de/SDSS/} 
have been excluded.

The physical properties of SDSS galaxies were analyzed in detail by \cite{Kauffmann03_masses,
Kauffmann03_sfh,Tremonti04} and \cite{B04} (hereafter B04) among others. 
In particular, the full likelihood distributions of their SFRs were derived by fitting all strong emission lines
simultaneously using the \cite{CharlotLonghetti01} models, following the methodology of 
\cite{Charlotetal02} (B04). Dust is accounted for with the \cite{CharlotFall00} multicomponent 
model which provides a consistent treatment of the attenuation of both continuum and emission line photons.
The dust attenuation is based on the H$\alpha$/H$\beta$ ratio to first order but is really constrained
by all the lines. B04 also devised a method for estimating the SFR of early-type galaxies with no 
detectable H$\alpha$ emission from their 4000\AA\ break index but we excluded those from our sample. 
We use the medians of the SFR distributions and consider these values, noted $SFR_e$ 
(for emission lines following B04), to be the best currently available estimates of the
SFR given the quality of the data and the technique used to derived them. Uncertainties 
are discussed in detail in the original paper. 

Other quantities such as 4000\AA\ break indices and stellar masses \citep{Kauffmann03_masses}
are also available from the SDSS MPA/JHU DR4 value added catalogs$^1$.
The stellar mass is defined as the total mass of stars formed over the lifetime of the galaxy. 
The 4000\AA\ break index -- D$_n$(4000) -- is defined as the ratio of the average flux density F$_\nu$ 
in the narrow bands 3850$-$3950\AA\ and 4000$-$4100\AA\ following \citet{Balogh98}. It is
a relatively dust insensitive measure of a galaxy's SFH, equivalent to the ratio of the 
SFR averaged over the last $\sim 10^{8.5}$ years to the SFR averaged over $>10^9$ years \citep{J07}.
The D$_n$(4000) distribution is strongly bimodal around D$_n$(4000)$\sim 1.6$, dividing galaxies into the well 
known `red sequence' of  early-type, old star dominated galaxies (D$_n$(4000)$\ga 1.6$) and the `blue cloud' 
of late-type galaxies with recent star formation (D$_n$(4000)$\la 1.6$) \citep{Kauffmann03_masses,Strateva01}. 
In the following we refer to these 2 populations as simply red and blue galaxies. 

Our final 2 samples consist of the 20800 galaxies in the primary sample (89\%) that have been detected
by GALEX in the NUV band (SDSS+NUV sample), and of the 14900 galaxies in the primary sample with additional
FUV coverage (85\%) that have been detected both in the NUV and FUV bands (SDSS+NUV+FUV sample). 
In the following we use the larger, SDSS+NUV only sample whenever FUV fluxes are not explicitly needed.
These samples are strongly biased against red sequence galaxies but complete 
for blue galaxies: $\sim 98\%$ of D$_n$(4000)$<1.6$ galaxies are detected in the NUV band 
(and in the FUV band when both are available) against $\sim 72\%$ of D$_n$(4000)$>1.6$ galaxies ($\sim 52\%$ 
in both UV  bands when both are available). The average magnitude error is 0.03 in the NUV band and 0.07 
in the FUV band for the blue population; 0.08 in the NUV band and 0.16 in the FUV band for the red population. 

We derive absolute magnitudes in all the bands from the redshift and the Galactic extinction-corrected 
SDSS+GALEX photometry using the {\tt kcorrect v4\_1} software of \citet{Blanton07_kcorrect}.
In order to minimize the uncertainties on the $k$-corrections, the magnitudes are $k$-corrected to the mean redshift 
of the SDSS sample ($\overline z=0.1$) and are noted $^{0.1}mag$ where $mag=f$ or $n$ for the GALEX FUV and NUV bands, 
$g$, $r$, $i$ and $z$ for the SDSS bands. The $k$-correction at redshift $\overline z$ is by definition $-2.5~{\rm log}(1+\overline z)$
in all bands for all galaxies and deviates from this value towards both ends of the redshift range ($0.005<z<0.28$).  
In the UV bands, this deviation is less than 0.1 magnitude for 95\% of the galaxies. 

\begin{figure}
%\plotone{cold4000.ps}
\plotone{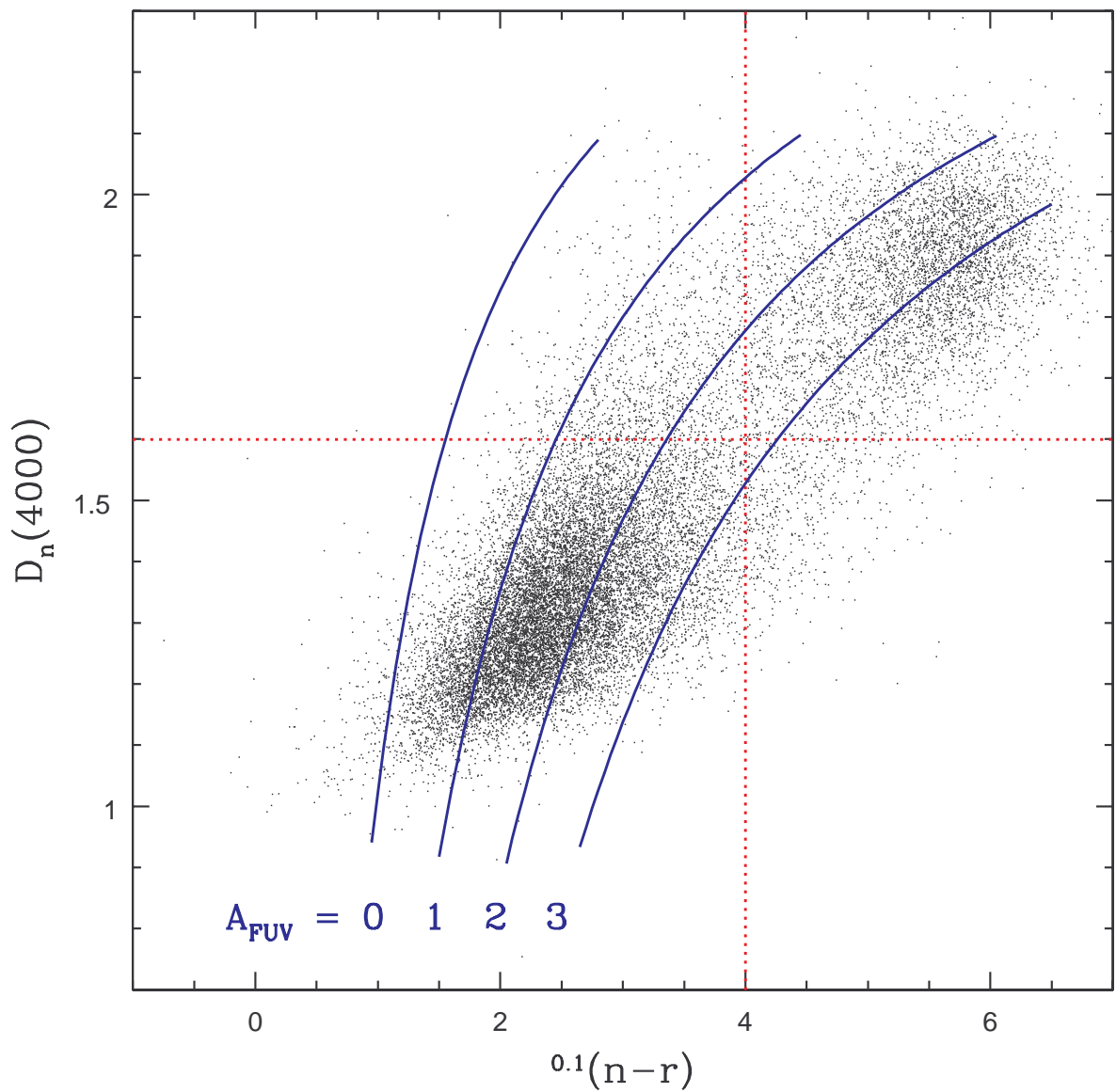}
\caption{The 4000\AA\ break index, D$_n$(4000), as a function of  $^{0.1}(n-r)$ for the SDSS+NUV sample. The blue curves 
represent lines of equal FUV attenuation by Johnson et al. (2007). The horizontal and vertical dotted lines at  D$_n$(4000)$=1.6$
and  $^{0.1}(n-r)=4$ mark the boundaries between the `red sequence' and `blue cloud' populations.
\label{fig:cold4000}}
\end{figure}

Figure \ref{fig:cold4000} shows the distribution of the SDSS+NUV sample in the $^{0.1}(n-r)$ 
vs D$_n$(4000) plane. The $^{0.1}(n-r)$ color distribution is strongly bimodal (Wyder et al. 2007)
with $^{0.1}(n-r)=4$ defining roughly the same boundaries as D$_n$(4000)$=1.6$ 
between the red and blue populations. Galaxies with $^{0.1}(n-r)<4$ and D$_n$(4000)$<1.6$ 
(the `blue cloud') represent 70\% of the sample (81\% of the SDSS+NUV+FUV sample); 
galaxies with $^{0.1}(n-r)>4$ and D$_n$(4000)$>1.6$  (the `red sequence') represent 22\% of the sample
(10\% of the SDSS+NUV+FUV sample). The solid lines are polynomial fits to the dust/color/SFH relation 
derived by \cite{J07}  (see Section 3) for given values of the FUV attenuation as marked in the figure.
The fits are good for galaxies with D$_n$(4000)$<1.6$ but less reliable for red  sequence galaxies (see Johnson et al. 2007 for details).
The model illustrates how broad band colors depend on both the SFH and the amount of dust  attenuation. 

\section{Deriving a Star Formation Rate from an Ultraviolet flux}

\subsection{Calibration}

The SFR measured from the UV emission is usually written as:
\begin{equation}
SFR_{UV}{\rm [M_\odot yr^{-1}]}={L_{UV}{\rm [erg~s^{-1}~Hz^{-1}]} \over \eta_{UV}}
\label{sfruv}
\end{equation}
where $\eta_{UV}$ converges to $\eta_{UV}^0$ for a constant SFR. Scaled to a Kroupa IMF, the most 
commonly used factor is log$(\eta_{UV}^0)=28.02$ \citep{Ken98}.  
It assumes that the UV spectrum is nearly flat in $L_\nu$ over 
the wavelength range 1500-2800 \AA. Using the \cite{BruzualCharlot03} stellar population
synthesis models with similar assumptions (solar metallicity, a constant SFH and a Kroupa IMF), 
S. Salim (private communication) derived slightly higher factors for the GALEX filters: 
log$(\eta_{FUV}^0)=28.09$ and log$(\eta_{NUV}^0)=28.08$. 
They are little sensitive to metallicity and to the SFH provided the SFR has been nearly constant 
in the last $10^{8}$ years. Very young starburst galaxies would significantly deviate from 
a constant SFR model and require a higher value of $\eta_{UV}$, while the UV emission of early-type 
galaxies is contaminated by older stars. For an optically selected sample with a mix of SFHs and 
metallicities similar to the SDSS/GALEX sample defined in the previous section (the average metallicity 
of which is $0.8 Z_\sun$), Samir et al. (2007) suggest using their median conversion factor 
log$(\overline \eta_{FUV})=28.14$. We assume this calibration for both UV bands in the following. 

\begin{figure*}
%\plotone{luv_sfr_0.ps}
\plottwo{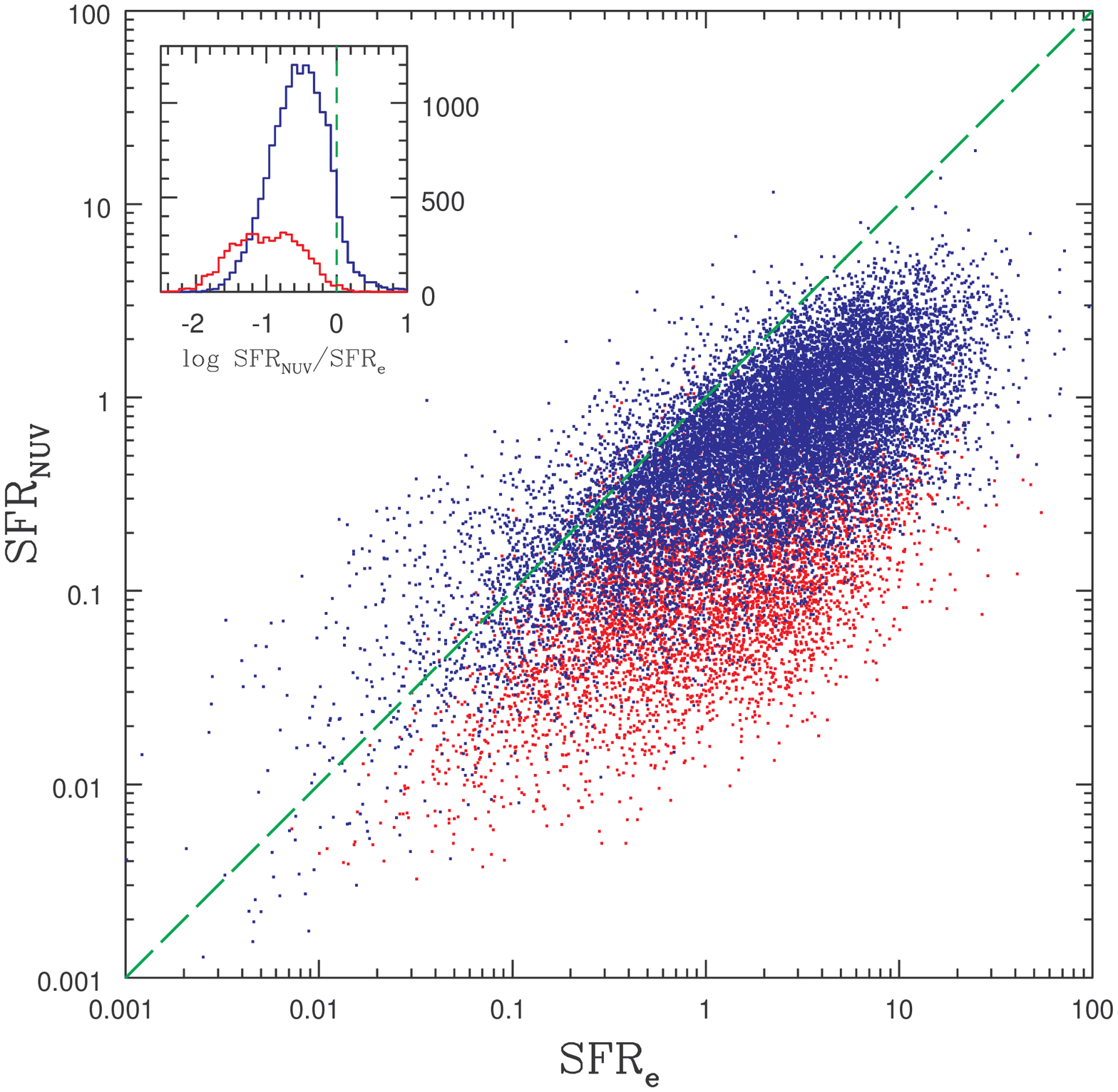}{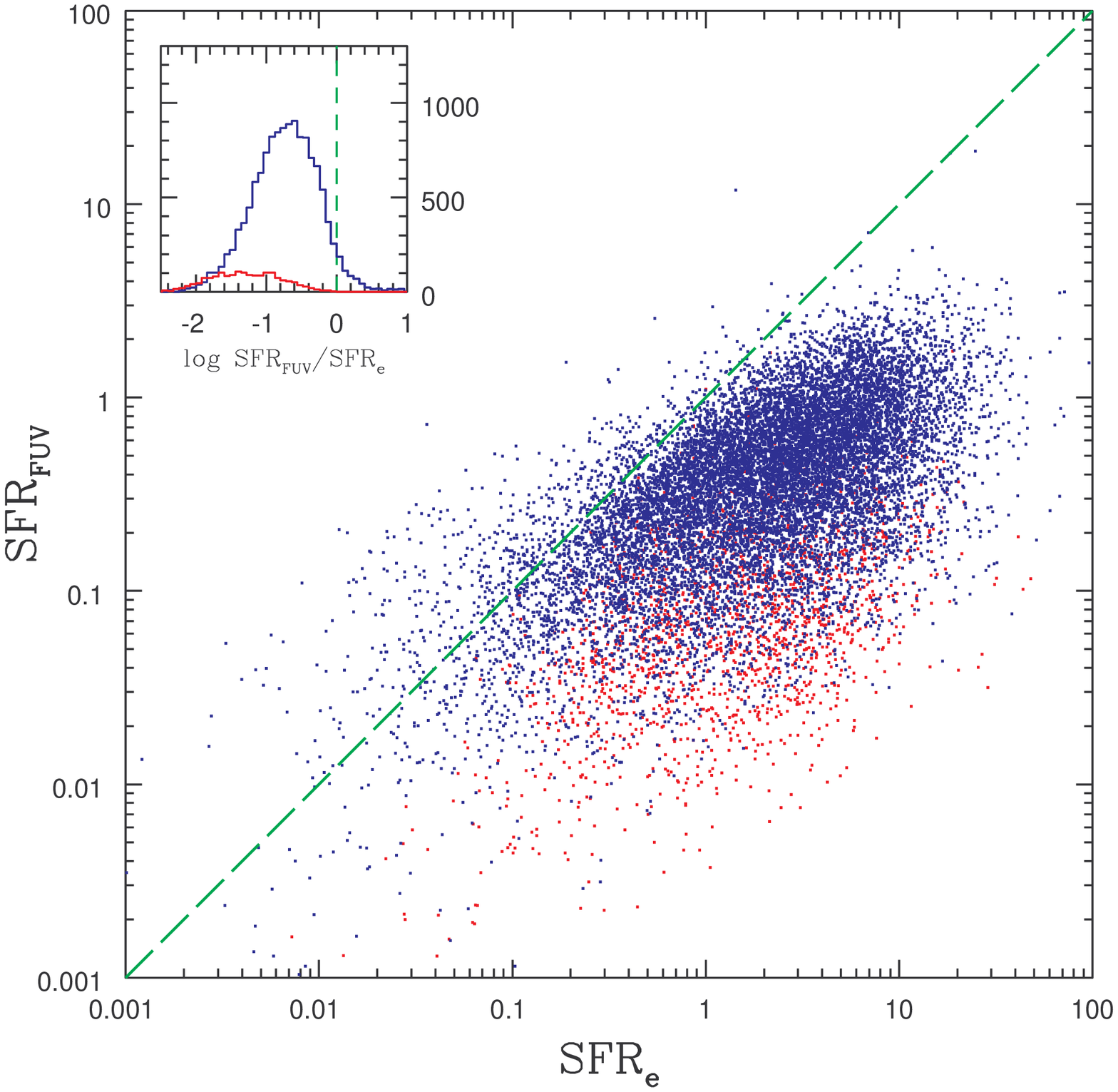}
\caption{The uncorrected UV SFRs ($SFR_{NUV}$ on the left; $SFR_{FUV}$ on the right) against the emission line derived $SFR_e$ 
\citep{B04} in units of M$_\odot~\rm yr^{-1}$. The dashed green lines shows SFR equality.
The blue and red dots represent blue and red sequence galaxies, defined as $^{0.1}(n-r)<4$ and $^{0.1}(n-r)>4$ respectively. 
The blue and red histograms in each panel are the ${\rm log}(SFR_{UV}/SFR_e)$ distributions for the two populations.
As expected from uncorrected luminosities $SFR_{UV}$ underestimates $SFR_e$,  especially so at high SFR.
\label{fig:luv_sfr_0}} 
\end{figure*}

Figure \ref{fig:luv_sfr_0} shows $SFR_{UV}$ against $SFR_e$ for the NUV and FUV
bands (left and right panel respectively), assuming no dust correction for the UV luminosities. 
The dashed green line denotes equality of SFR. Blue and red dots distinguish between blue and red galaxies 
defined as having $^{0.1}(n-r)<4$ and $>4$ respectively. The histograms in inset show the distribution of the
$SFR_{UV}$ to  $SFR_e$ ratios for the blue
and red populations.  As expected from uncorrected luminosities, $SFR_{UV}$ underestimates the `true' SFR,
including for the red population although part of their UV luminosity is unrelated to the current SFR.
The scatter is large in both bands, indicating a large range of UV attenuations for a given SFR or a given UV luminosity. 
There is also a clear trend with $SFR_e$ in the sense that galaxies with higher SFR tend to require a larger dust correction,
as was first observed by \citet{WangHeckman96}. 

\subsection{Dust attenuation estimates}
 
The fraction of UV flux emitted by new born stars and absorbed by dust in the galaxy is reradiated at infrared (IR) 
wavelengths \citep{Buat92}. Assuming a standard extinction law and that the dust is heated by intrinsically young 
stellar populations, the FUV attenuation can be approximated by:
\begin{equation}
A_{FUV}=2.5\times {\rm log}(\mu IRX+1)
\label{meurer}
\end{equation}  
where IRX is the ratio of the IR to UV luminosities (the so called infrared excess) and
$\mu$ corrects for the fraction of IR luminosity heated by older stars and by light
bluer than the FUV band \citep{Meurer99}. Other relations were derived that yield 
very similar results (e.g. Buat et al. 2005\nocite{Buat05}).
UV reddening as measured by the slope $\beta$ of the 
UV continuum ($f_\lambda \propto \lambda^{\beta}$) or a UV color, correlates with IRX 
in starburst galaxies, as expected from a foreground screen of dust \citep{Witt92,Calzetti94}. 
Thus $\beta$ or UV colors are often used to estimate $A_{FUV}$. 
The IRX/$\beta$ correlation was recently revisited and corrected by several authors for more 
`normal' star-forming galaxies using GALEX data 
(Seibert et al. 2005, Cortese et al. 2006, Salim et al. 2007, Johnson et al. 2007). 
\cite{Se05} (hereafter Se05) found that the starburst relation systematically overestimates
the FUV attenuation of more quiet galaxy types by 0.58 mag, albeit with a large scatter. They
derived the following empirical relation from a diverse sample of $\sim 200$ galaxies with UV 
photometry from GALEX and FIR photometry from the {\it Infrared Astronomical Satellite} ({\it IRAS}):
\begin{equation}
A_{FUV}^{Se05}=3.97~(m_{FUV}-m_{NUV}-0.1)+0.14
\label{mark}
\end{equation}
where $m_{FUV}$ and $m_{NUV}$ are the apparent magnitudes in the FUV and NUV bands respectively. 
The 0.1 magnitude offset corrects for a change in calibration between the GALEX photometry used by Se05 
(the internal data release IR0.2) and that used in the present paper (IR1.1) (Seibert et al., in
preparation).

\cite{Sa07} (hereafter Sa07) derived a yet shallower relation between the attenuation 
and the UV color of normal blue galaxies using a different technique and the much larger 
GALEX/SDSS sample we are using here.
They obtained the SFR and FUV attenuations, among other physical properties, by fitting the
UV and optical photometry to an extensive library of model SEDs for which dust attenuation was
computed from the same \cite{CharlotFall00} model used by B04. While B04 obtained their SFRs 
and attenuations from optical emission lines (the H$\alpha$ line and the Balmer decrement to first 
approximation), Sa07's are essentially constrained by the UV fluxes and the UV colors respectively. 
The agreement is generally good between the two approaches but discrepancies remain, in particular 
between the attenuations as we'll discuss further below. Sa07 derived the following simple prescription
from their extensive modeling for galaxies with $^0(n-r)<4$ and $^0(f-n)<0.9$:
\begin{equation}
A_{FUV}^{Sa07}=2.99~ ^0(f-n)+0.27 
\label{samir}
\end{equation}
where the 0 subscript refers to rest-frame colors $k$-corrected to $z=0$. The small ($<5\%$) fraction of
galaxies with $^0(n-r)<4$ and $^0(f-n)>0.9$ are assigned a constant attenuation of 2.96.

Longer baseline colors such as UV$-$optical colors carry mixed but separable information about the 
SFH and IRX. Using a sample of galaxies with UV through IR photometry from GALEX, SDSS and Spitzer, 
Johnson et al. (2006, 2007) (hereafter J07)\nocite{J06} showed that given the SFH of a galaxy 
(they used D$_n$(4000)), IRX could be more accurately inferred from UV$-$optical colors than from UV colors. 
Assuming Eq. \ref{meurer} with $\mu=0.6$, they derived the following relation for galaxies with  D$_n$(4000)$<1.6$:
\begin{equation}
A_{FUV}^{J07}=1.21 -2.04 x +1.45 y -0.98 y^2
\label{ben}
\end{equation} 
where $x=D_n(4000)-1.25$ and $y=~^{0.1}(n-r) -2$. 

\begin{figure*} 
%\plotone{luv_sfr_all.ps}
\plotone{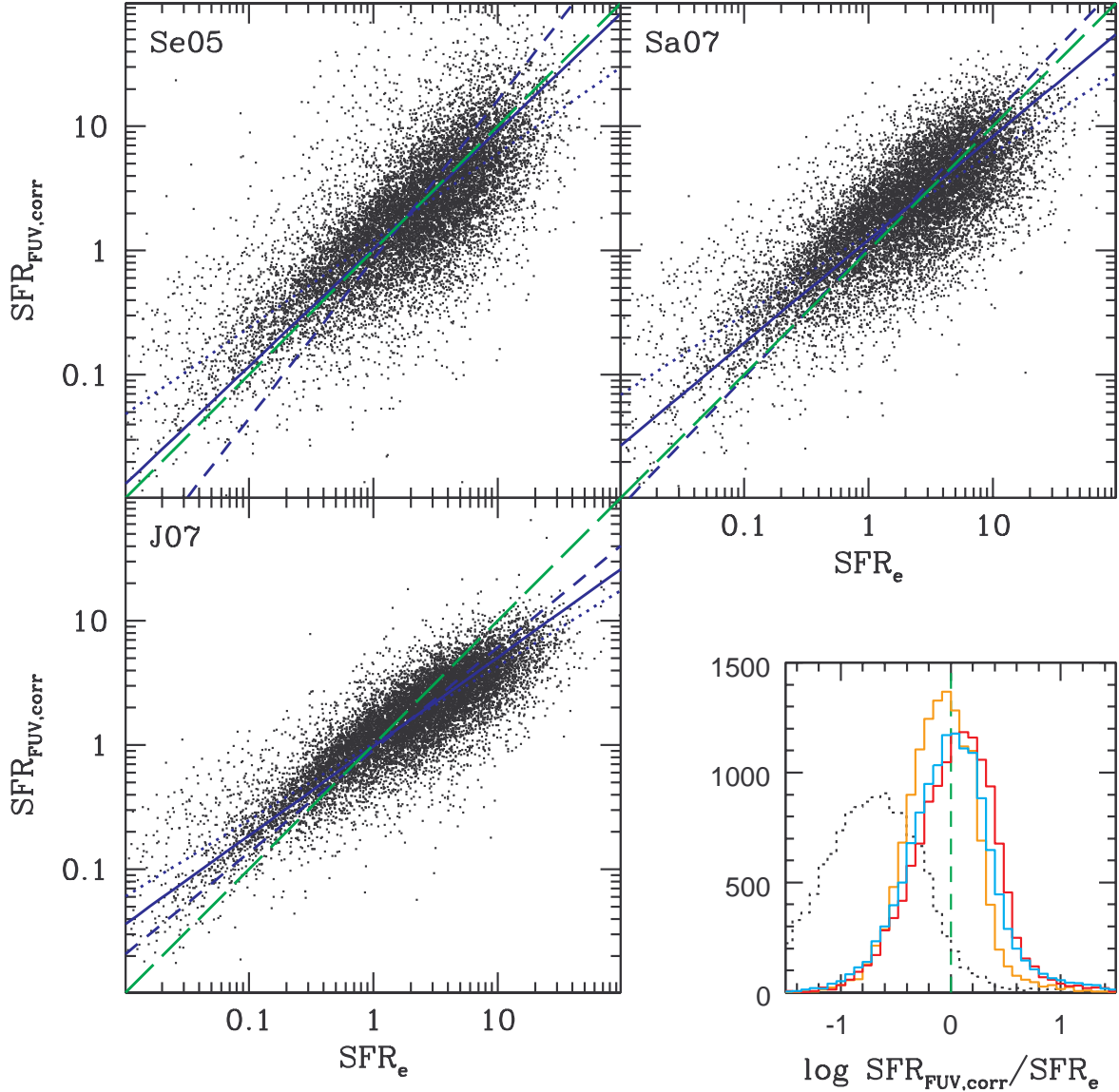}
\caption{The dust corrected FUV SFRs ($SFR_{FUV,corr}$) against $SFR_e$ for the blue population ($^{0.1}(n-r)<4$) 
using the FUV attenuation methods of
\citet{Se05} (Se05), \citet{Sa07} (Sa07) and \citet{J07} (J07)  as indicated. The SFRs are  in units of M$_\odot~\rm yr^{-1}$.
The dotted, dashed and solid 
blue lines in each panel are the ordinary least-square (OLS) regression of the Y axis 
on the X axis, the OLS regression of the X axis on the Y axis and the bisector of those 2 lines respectively. 
The histograms show the distributions of the $SFR_{FUV,corr}/SFR_e$ logarithmic ratios: light blue, red and orange
for Se05, Sa07 and J07 respectively. The dotted histogram is the distribution of the uncorrected $SFR_{FUV}/SFR_e$
ratios. All three models provide a very good average correction but tend to over (under) correct galaxies with
the lowest (highest) SFRs.
\label{fig:luv_sfr_all}
}
\end{figure*}

We note $SFR_{FUV,corr}$ the SFR derived from the UV luminosity corrected for dust attenuation using one 
of the above equations. 
Figure \ref{fig:luv_sfr_all} shows  $SFR_{FUV,corr}$ against $SFR_e$ for the blue galaxies ($^{0.1}(n-r)<4$)
using  Eq. \ref{mark} (Se05), \ref{samir} (Sa07) and \ref{ben} (J07) as indicated. In each panel
the dotted, dashed and solid blue lines show the ordinary least-square (OLS) regression of the Y axis 
on the X axis, the OLS regression of the X axis on the Y axis and the bisector of those
2 lines respectively \citep{Isobe90}. We choose the bisectors as the `best-fit' lines, here and 
in the rest of the paper. The best-fit slopes, variances, correlation coefficients and 
residual scatters are listed in Table 1 for the 3 attenuation models (first 3 lines).
The histograms show the distributions of the $SFR_{FUV,corr}/SFR_e$ ratios compared to
the distribution of the uncorrected $SFR_{FUV}$ to $SFR_e$ ratios. The averages  of the
distributions are also listed in Table 1. 
All 3 methods provide a very good average correction with a reduced scatter compared to the
uncorrected $SFR_{FUV}$, especially so for the J07 correction. 
However a residual trend with $SFR_e$ remains in the sense that galaxies with the highest and 
lowest SFR tend to be under and over-corrected respectively. This indicates that the models
do not quite sufficiently scale with the SFR to straighten up the uncorrected correlation in 
Fig. \ref{fig:luv_sfr_0}.  The trend is minimal for the Se05 correction for which the scatter is largest
and most pronounced for the J07 correction for which the scatter is otherwise best reduced. 
This trend with $SFR_e$ is the same as that noted by Sa07 as a trend with mass 
(their Fig. 8) and by J07 as a trend with $^{0.1}(n-r)$ and D$_n$(4000) (their Fig. 12 and 13). 
Indeed mass and to a lesser extent colors and D$_n$(4000) correlate with $SFR_e$. The reason for
it remains unclear but Sa07 concluded that the most likely interpretation in the framework of their 
modeling was that attenuations were less well constrained by the UV data than by the emission lines 
at the two ends of the distribution. 
In the case of J07, the parametric relation between IRX, D$_n$(4000) and $^{0.1}(n-r)$ is a good fit to 
blue galaxies, more so than between IRX and UV color (see J07 for a detailed discussion).
Therefore it is perhaps the relation between IRX and $A_{FUV}$ that is not totally adequate. 
We return to this point in the next section.

\section{Reconciling UV and emission line Star Formation Rates}

\subsection{Empirical color corrections}

\begin{figure*}
%\plottwo{luv_sfr_col_n.ps}{luv_sfr_col_f.ps}
\plottwo{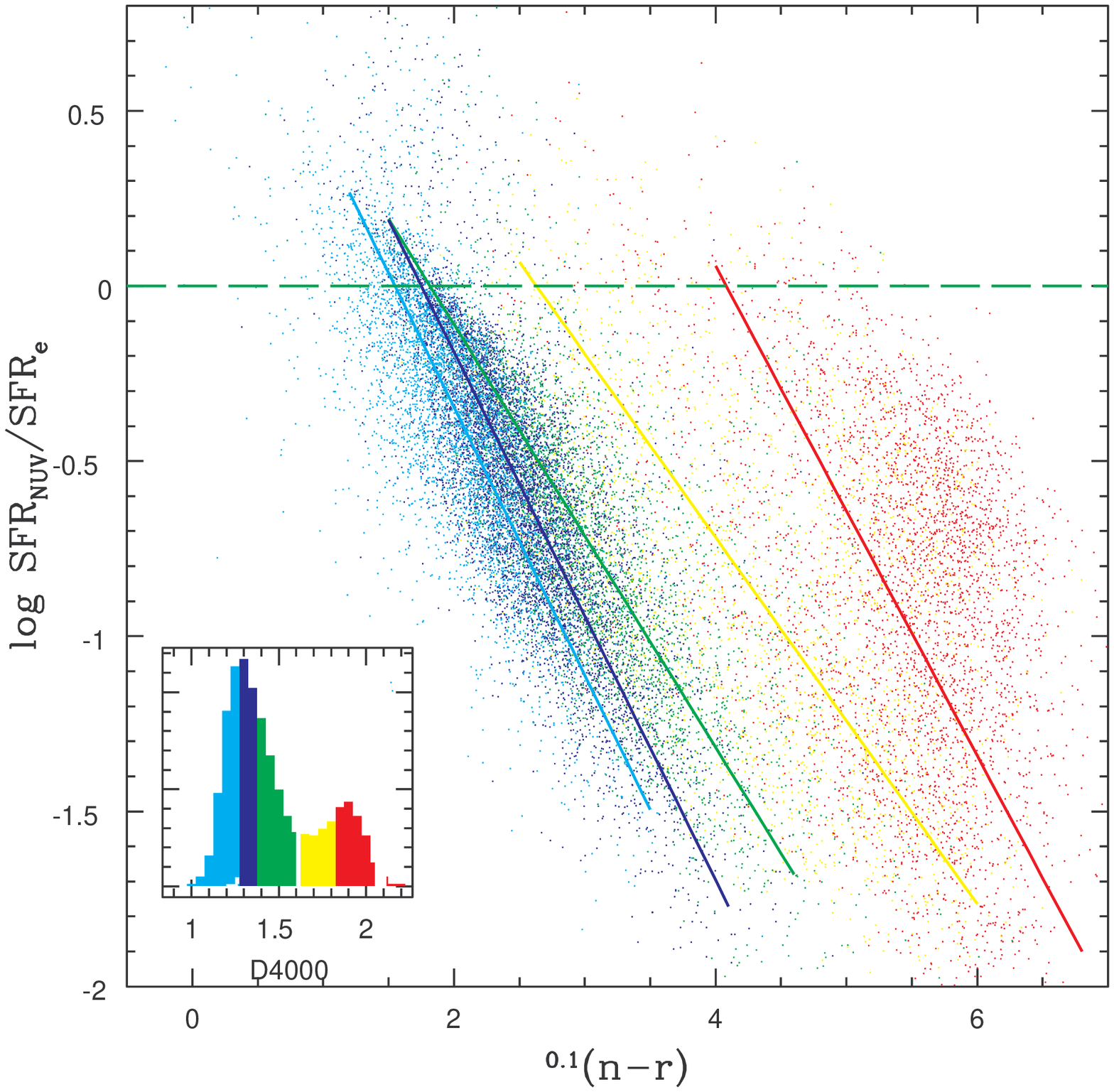}{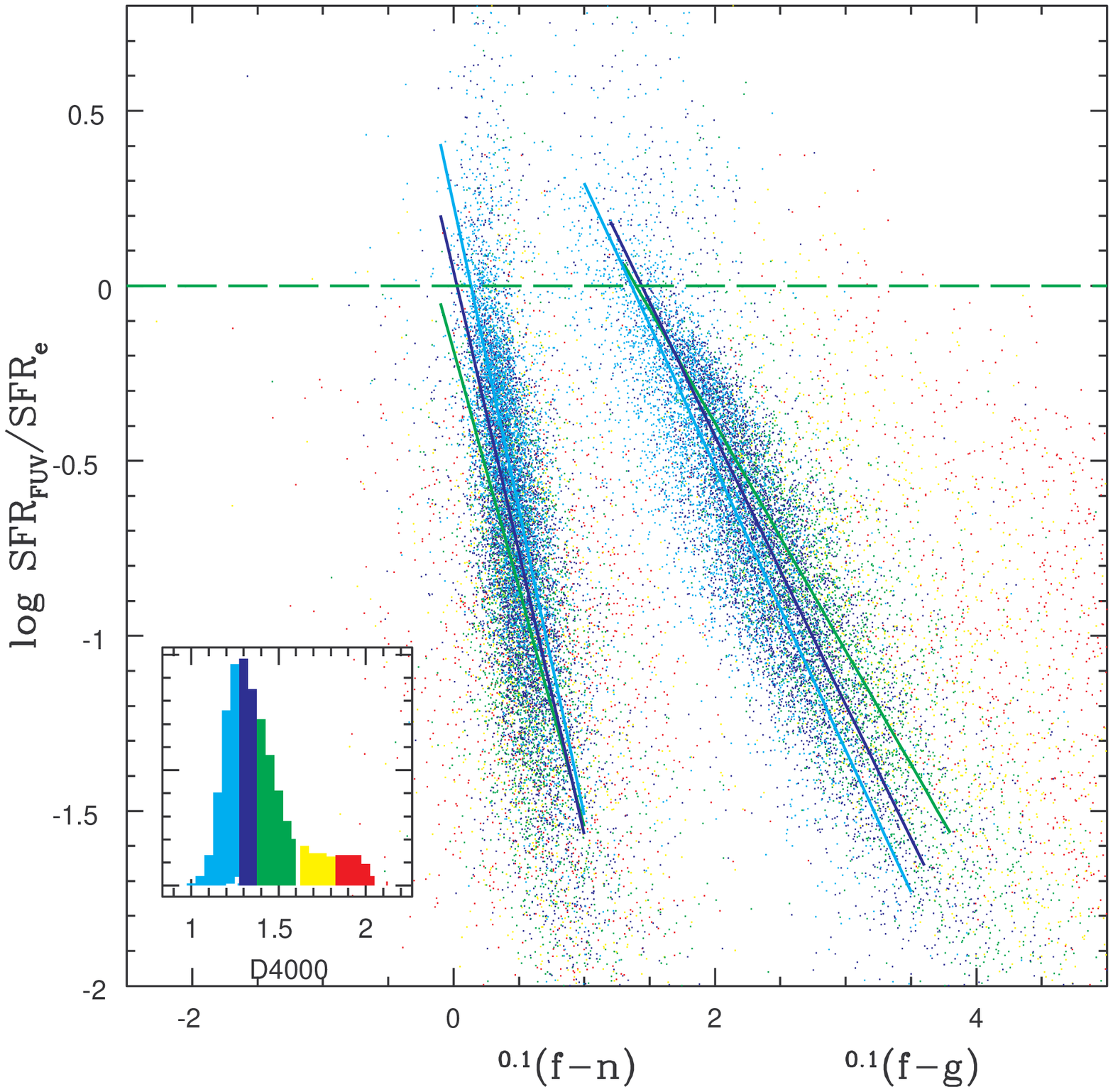}
\caption{The color dependence of the $SFR_{UV}$ to $SFR_e$ ratio for different bins of D$_n$(4000).
Left: ${\rm log} (SFR_{NUV}/SFR_e)$ against $^{0.1}(n-r)$ for 
the SDSS+NUV sample;  Right: ${\rm log}(SFR_{FUV}/SFR_e)$ against $^{0.1}(f-g)$ and against $^{0.1}(f-n)$ for the SDSS+NUV+FUV sample. 
The data are color coded to match the binned D$_n$(4000) distribution histogram shown in inset. 
The colored solid lines show the fitted linear correlations in each bin of D$_n$(4000). The 3 bins making up the  `blue cloud'
population (D$_n$(4000)$<1.6$: light blue, dark blue and green) are tightly correlated with color, while the 2 bins forming the red sequence 
(D$_n$(4000)$>1.6$: yellow and red) are practically independent of it.  The color correlations observed for blue galaxies provide simple UV
attenuation corrections (see text for details). 
\label{fig:luv_sfr_col}}
\end{figure*}

Assuming as we do that $SFR_e$ is the current best dust-corrected SFR estimate and that our choice of
$\eta_{UV}$ is adequate, the UV attenuation (FUV or NUV) can be directly measured as: 
\begin{eqnarray}
A_{UV}=-2.5~ {\rm log}(SFR_{UV}/SFR_e)
\label{attenuation}
\end{eqnarray}
We now revisit the color dependence of these known attenuations. 
Figure \ref{fig:luv_sfr_col}, left panel, shows the $SFR_{NUV}$ to $SFR_e$ ratios as a function 
of $^{0.1}(n-r)$ for the SDSS+NUV sample. The solid lines show the OLS bisector for each D$_n$(4000) 
bin as shown in inset. The 3 bins making up the blue sequence (D$_n$(4000)$<1.6$) add up to form a single tight 
correlation while the 2 bins with D$_n$(4000)$>1.6$ form a scattered cloud. 
For galaxies with D$_n$(4000)$<1.6$ and $^{0.1}(n-r)<4$, the bisector fit is:
\begin{equation}
A_{NUV,n-r} = 1.71~ ^{0.1}(n-r) -2.86
\label{Anuv_nr}
\end{equation}
with a linear correlation coefficient $r=0.77$ and $rms=0.70$.
The SDSS+NUV sample allows us to see the impact of the
SFH on the attenuation/color relation. It is bimodal to first order: attenuation is linearly
dependent on color for blue sequence galaxies and practically independent of it for red sequence
galaxies. Similar correlations are found when using FUV luminosities and/or other UV$-$optical
colors. Using the FUV luminosity and $^{0.1}(f-g)$ yields:
\begin{equation}
A_{FUV,f-g} = 1.84~^{0.1}(f-g) -2.57.
\label{Afuv_fg}
\end{equation}
The correlation is tighter than Eq. \ref{Anuv_nr} ($r=0.84$, $rms=0.66$). On the other hand $^{0.1}(f-n)$
results in a poorer and more scattered correlation ($r=0.56$, $rms=1.02$):
\begin{equation}
A_{FUV,f-n} = 4.05~^{0.1}(f-n) -0.18.
\label{Afuv_fn}
\end{equation}
The last two correlations are shown in the right panel of Fig.~\ref{fig:luv_sfr_col}.
The latter correlation is similar to that proposed by \cite{Cortese06} (their Fig. 10) 
using log$(L_{\rm H\alpha}/L_{FUV})$ as a function of $(m_{FUV}-m_{NUV})$ for a small sample of star-forming 
galaxies in the COMA cluster.
It is consistent with Eq. \ref{mark} (Se05) but a much steeper function of UV  color than Eq. \ref{samir} (Sa07).
However all 3 equations as well as Eq. \ref{Afuv_fg} converge for the majority of galaxies around 
the peak of the color distribution ($^{0.1}(n-r)\sim 2.4-2.5$) and yield similar average values 
in good agreement with the measured average attenuation ($<A_{FUV}>\sim 1.8$).
Equation \ref{ben} (J07) yields a slightly lower average attenuation of 1.6. 
Discrepancies between these corrections are largest for the reddest and bluest galaxies. 

\begin{figure*} 
%\plottwo{luv_sfr_4.ps}{luv_sfr_3.ps}
\plottwo{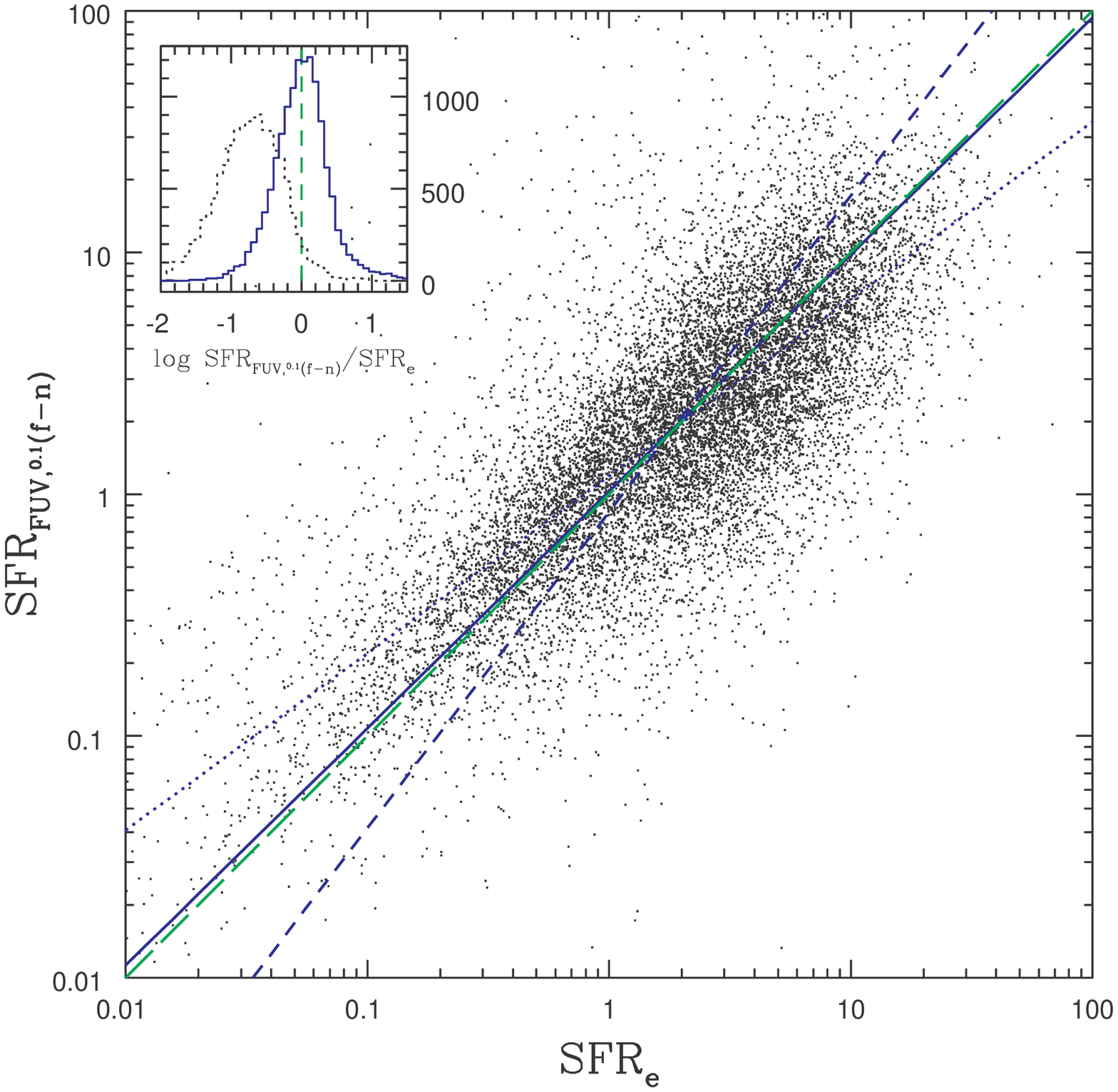}{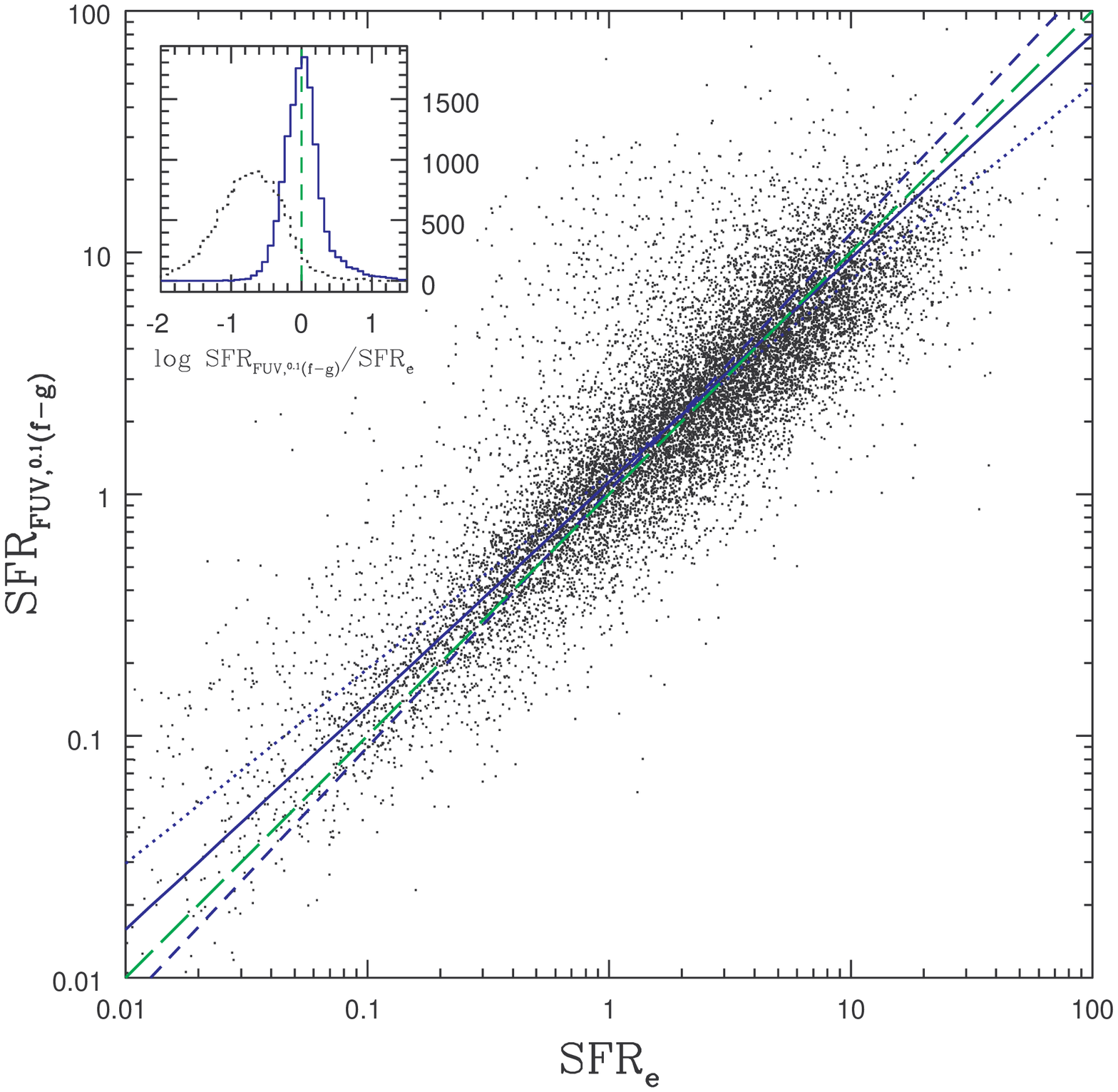}
\caption{The color corrected FUV SFRs against $SFR_e$. The FUV luminosities are corrected using
the correlations established in Fig.~\ref{fig:luv_sfr_col} between $A_{FUV}$ and $^{0.1}(f-n)$ (Eq.~\ref{Afuv_fn}; left panel) and between
$A_{FUV}$ and $^{0.1}(f-g)$ (Eq.~\ref{Afuv_fg}; right panel). 
The histograms in inset show the distributions of ${\rm log}(SFR_{FUV}/SFR_e)$ with and without corrections (solid and dotted lines respectively). 
$SFR_e$ is best recovered using the $A_{FUV}$/$^{0.1}(f-g)$ correlation. 
\label{fig:luv_sfr_c}
}
\end{figure*}

\begin{figure} 
%\plotone{comp_corr_all.ps}
\plotone{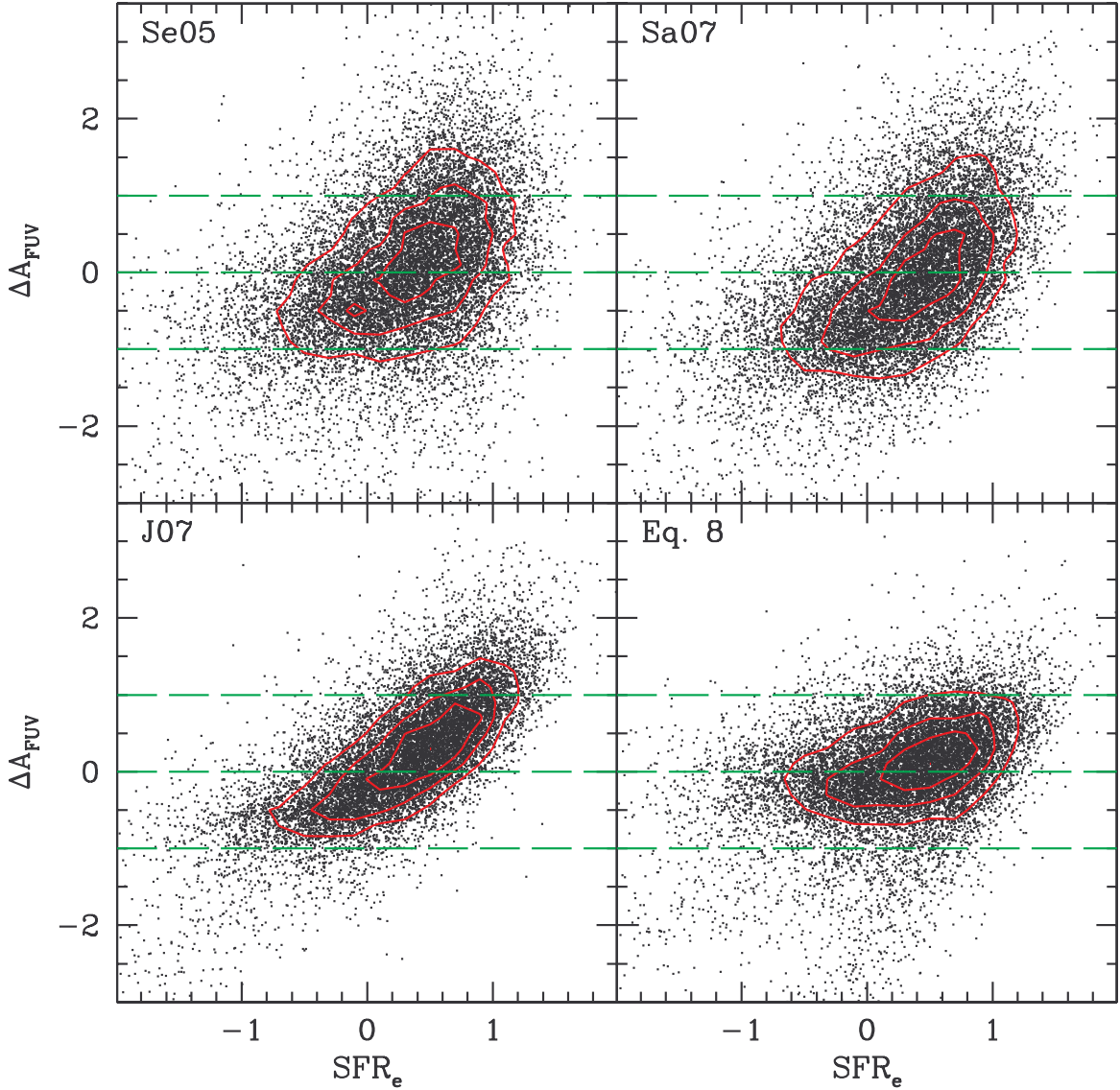}
\caption{$\Delta A_{FUV}= A_{FUV}-A_{FUV,{\rm model}}$ as a function of $SFR_e$ for the various models described in the text as indicated
(Sa05, Se07, J07 and Eq. \ref{Afuv_fg}). The dashed lines mark $\Delta A_{FUV}=-1$, 0 and +1.  The red curves are isodensity contours. 
All 4 methods converge with $A_{FUV}$ around the peak of the SFR distribution ($\sim 2 {\rm M}_{\odot} {\rm yr}^{-1}$ ) and provide good 
average corrections but Eq. \ref{Afuv_fg} (the $A_{FUV}$/$^{0.1}(f-g)$ correlation) minimizes $\Delta A_{FUV}$ and its dependence with 
$SFR_e$ for the majority of the galaxies.
\label{fig:comp_corr_all}}
\end{figure}

We note $SFR_{FUV,c}$ the FUV based SFRs corrected using one of the functions of color derived in 
this section.  Figure \ref{fig:luv_sfr_c} shows $SFR_{FUV,c}$ against $SFR_e$ using $A_{FUV,f-n}$ (left panel) 
and $A_{FUV,f-g}$ (right panel) for the blue population.
The correlation is very close to equality in both cases, as expected since the corrections were 
designed to minimize $SFR_{FUV,c}/SFR_e$, and the scatter is best reduced using $^{0.1}(f-g)$, as 
expected as well from the higher correlation coefficient in Eq. \ref{Afuv_fg}. However a small residual
trend with $SFR_e$ remains in this case, which means that the SFR dependence  of the attenuation
is not completely accounted for by the color dependence (${\rm log}(SFR_{UV}/SFR_e$) would be better
fitted by a linear function of color plus a linear function of ${\rm log}(SFR_e)$). The trend practically disappears
when using $^{0.1}(f-n)$ but the correlation is less significant, as in the case of Se05. 
The parameters of the fits are listed in Table 1 for comparison
with the corrections presented in the previous section.  There is a small trade-off between the scatter 
and the trend with $SFR_e$ (the lower the rms, the more the slope deviates from unity) except for the empirical 
$^{0.1}(f-g)$ correction for which the combination of trend and scatter is best reduced ($a>0.9$ and $rms<0.3$). 
Figure \ref{fig:comp_corr_all} shows the difference between the measured attenuation 
(Eq. \ref{attenuation}) and the 4 parametric estimates (Se05, Sa07, J07 and Eq. \ref{Afuv_fg}) 
as a function of $SFR_e$ ($\Delta A_{FUV}= A_{FUV}-A_{FUV, {\rm model}}$).
The red curves are isodensity contours. The dashed green lines mark $\Delta A_{FUV}=-1$, 0 and 1. 
All 4 methods converge with $A_{FUV}$ around the peak of the SFR distribution ($\sim 2 {\rm M}_{\odot} {\rm yr}^{-1}$ ) 
and as noted above provide good average corrections but Eq. \ref{Afuv_fg} minimizes $\Delta A_{FUV}$ 
for the majority of the galaxies as well as the dependence with $SFR_e$.
Although the improvement is by no means dramatic, it provides an estimate of 
the FUV attenuation which best recovers the SFR derived from emission lines at an equivalent 
or lower `cost', since at least one optical photometric measurement and the redshift are required 
for $k$-corrections in all cases.

\subsection{Discussion}

\begin{figure}
%\plotone{luv_sfr_ap.ps}
\plotone{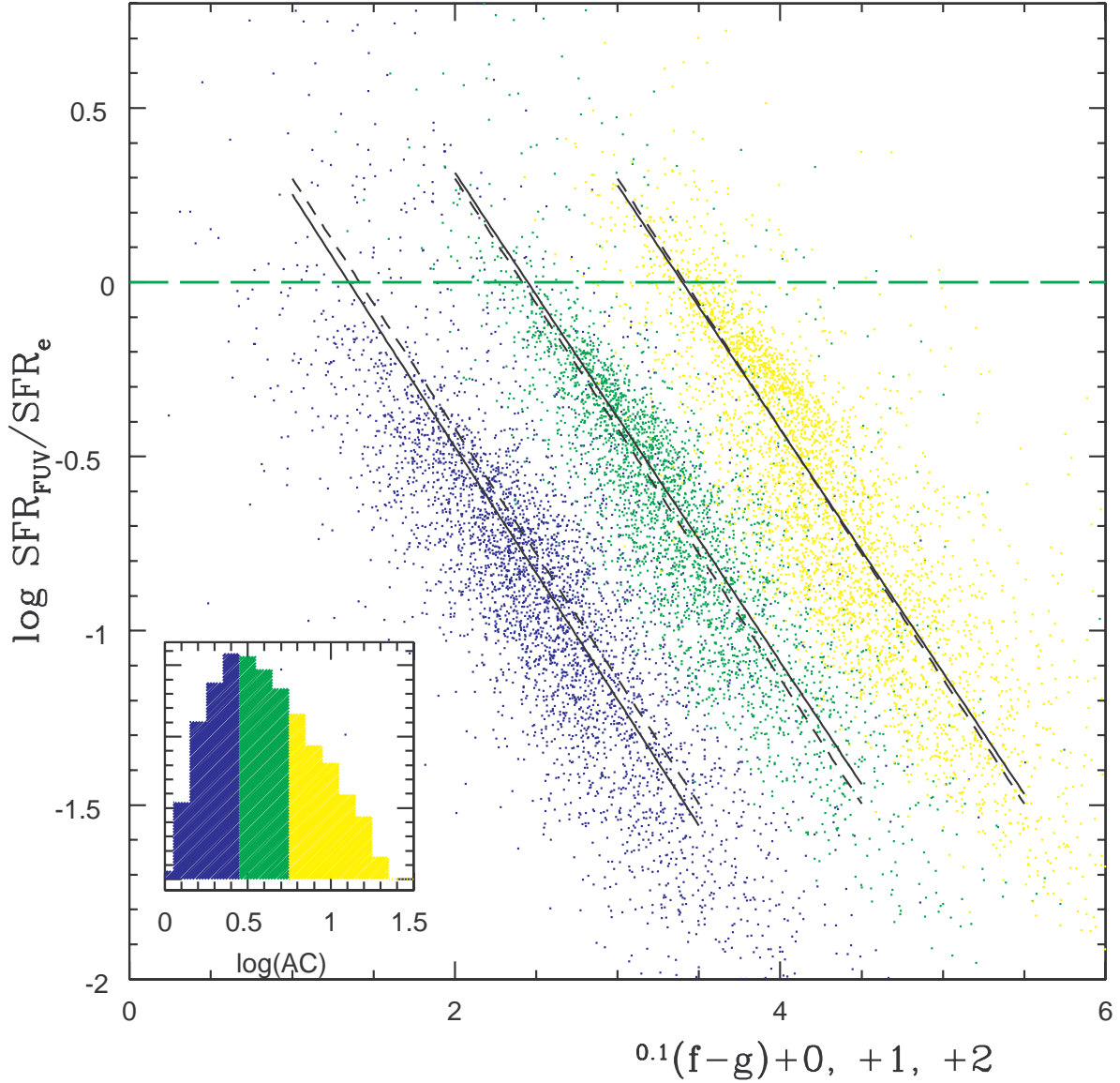}
\caption{The $SFR_{FUV}$ to $SFR_e$ ratio as a function of $^{0.1}(f-g)$ in 3 bins of aperture corrections (AC) 
as shown in inset. The AC is defined as the ratio of the total SFR ($SFR_e$) to the SFR within the SDSS fiber. 
For clarity the data are shifted by 0, +1 and +2 for the first, second and third bin respectively. The solid lines are the 
linear fits in each bin while the dashed lines show the correlation for the full sample (Eq. \ref{Afuv_fg})
shifted by the appropriate amount. Aperture effects do not bias the correlation.
\label{fig:luv_sfr_ap}}
\end{figure}

As aperture corrections are an important source of uncertainty in deriving total SFRs from the
SDSS fiber spectra, we check that the above correlations are not affected by aperture
effects. We define the aperture correction (AC) as the ratio of the total SFR ($SFR_e$) to the
SFR estimated within the fiber (B04) and split 
the blue sample into 3 bins of increasing aperture correction. Figure \ref{fig:luv_sfr_ap} 
shows the $SFR_{FUV}$ to $SFR_e$ ratios as a function of $^{0.1}(f-g)$ for the 3 bins
(shown in inset), shifted by a constant as indicated for clarity. The solid lines show
the bisector fits in each bin and the dashed lines show Eq. \ref{Afuv_fg} shifted by the 
appropriate amount for comparison. No significant difference is seen in the correlation itself 
as a function of aperture correction,
but galaxies with $SFR_{FUV}>SFR_e$ (yielding a negative FUV attenuation) have systematically 
high aperture corrections (and low redshift). It is likely that in these large nearby galaxies, 
the fiber missed off-centered regions of enhanced star-formation and that $SFR_e$ was underestimated.

\begin{figure}
%\plotone{Anuv_Afuv.ps}
\plotone{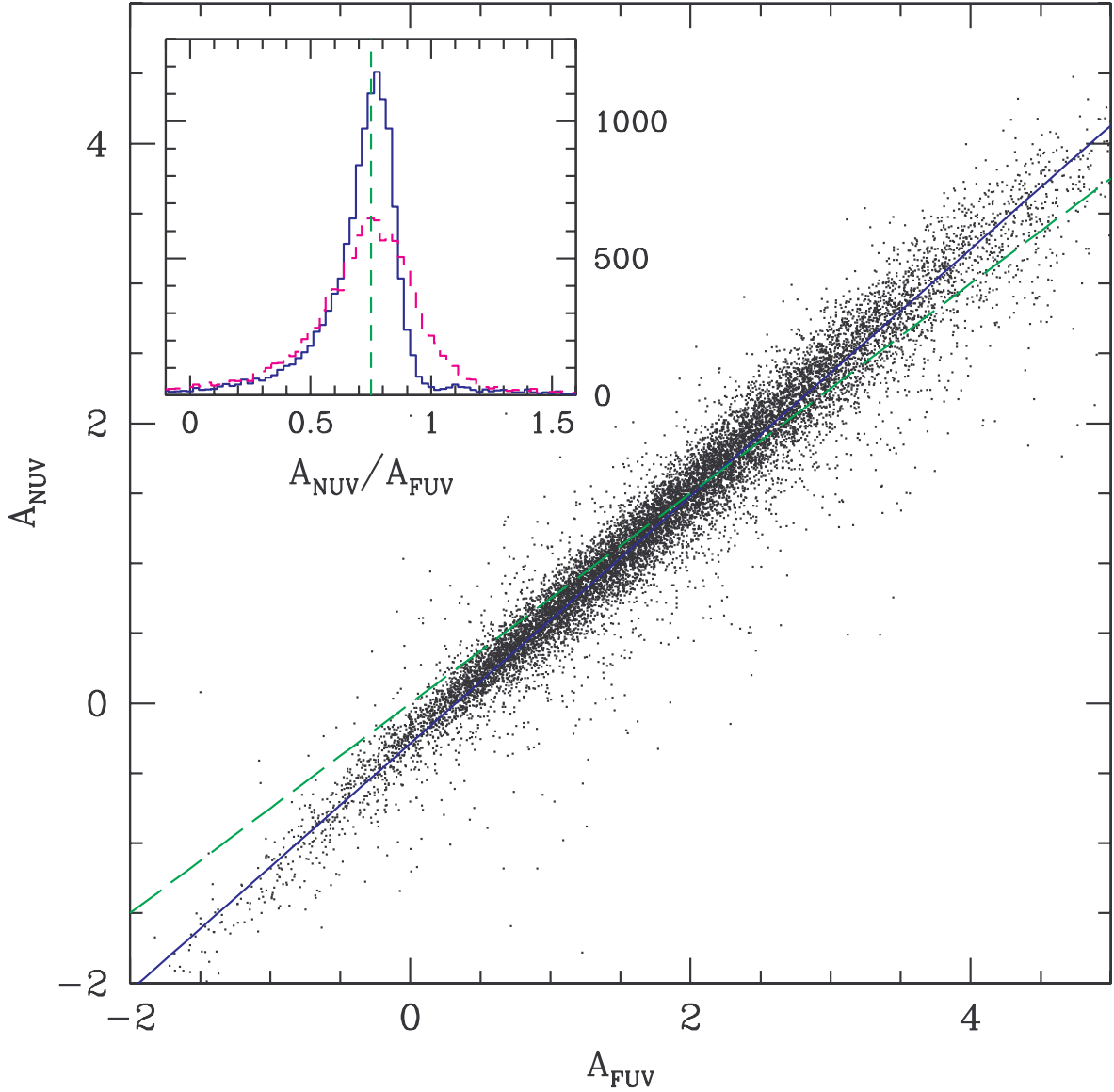}
\caption{The relation between $A_{NUV}$ and $A_{FUV}$ as defined by the $SFR_{UV}/SFR_e$ ratios (Eq. \ref{attenuation}). 
The solid blue line is the fitted correlation;  the dashed green line at $A_{NUV}=0.75~A_{FUV}$ is the ratio expected from a 
$\lambda^{-0.7}$ absorption curve (Charlot \& Fall 2000). The distribution of the $A_{NUV}$ to $A_{FUV}$ ratios is shown in inset 
(solid histogram) with that of the fits ratios ($A_{NUV,n-r}/A_{FUV,f-g}$; dotted histogram). The median ratio for the measured 
quantities is 0.74 (0.75 for the fits), in excellent agreement with the prediction. 
\label{fig:Anuv_Afuv}}
\end{figure}

Another source of uncertainty may lie in the definition of the UV attenuation (Eq.~\ref{attenuation}).
Allowing $\eta_{UV}$ to vary, the $SFR_{UV}$ to $SFR_e$ ratio would no longer be a measure of 
attenuation alone but a combination of  $A_{FUV}$ and $\eta_{UV}$: 
log$(SFR_{UV}/SFR_e)=-0.4A_{UV}+{\rm log}(\eta_{UV}/\overline \eta_{UV})$. 
B04 showed that the conversion factor $\eta_{\rm H\alpha}$ from H$\alpha$ luminosity to SFR 
($\eta_{\rm H\alpha}=L_{\rm H\alpha}/SFR_e$) decreases with mass (and metallicity), spanning
nearly 0.4 dex in the range of mass (and metallicity) spanned by the SDSS sample. This is interpreted 
as massive/metal rich galaxies producing less H$\alpha$ than low mass, metal poor galaxies for the 
same SFR. However $\eta_{UV}$ is much less sensitive to metallicity than H$\alpha$, 
consequently $SFR_{UV}/SFR_e$ is expected to be a nearly direct measure of UV attenuation except 
perhaps for galaxies with very low UV attenuation and metallicity. In any case the $\sim 1$ dex variation in $\eta_{UV}$ 
required to straighten up the trend seen in Fig. \ref{fig:comp_corr_all}, most notably for the J07 dust correction, is definitely ruled out. 
Let's note that Eq. \ref{Afuv_fg} can be used to recover $SFR_e$ from the observed UV luminosity
whatever the interpretation of the $SFR_{UV}/SFR_e$ ratio provided $\overline \eta_{UV}$ is assumed 
in Eq. \ref{sfruv}. 
Figure \ref{fig:Anuv_Afuv} shows the relation between $A_{NUV}$ and $A_{FUV}$ as defined in Eq. \ref{attenuation}. 
The solid blue line is the best linear fit; 
the dashed green line at $A_{NUV}=0.75~A_{FUV}$ is the ratio expected from a $\lambda^{-0.7}$ 
absorption curve \citep{CharlotFall00}.
The distribution of the $A_{NUV}$ to $A_{FUV}$ ratios is shown in inset as a solid histogram, and 
that of the $A_{NUV,n-r}$ to $A_{FUV,f-g}$ ratios as a dotted histogram. 
The median ratio for the measured quantities is 0.74 (0.75 for the fits), in excellent agreement with 
the prediction. The GALEX data are therefore consistent with the attenuated UV fluxes predicted by
the dust model used to derive $SFR_e$ from independent emission line measurements. 

\begin{figure} 
%\plotone{irx_Afuv.ps}
\plotone{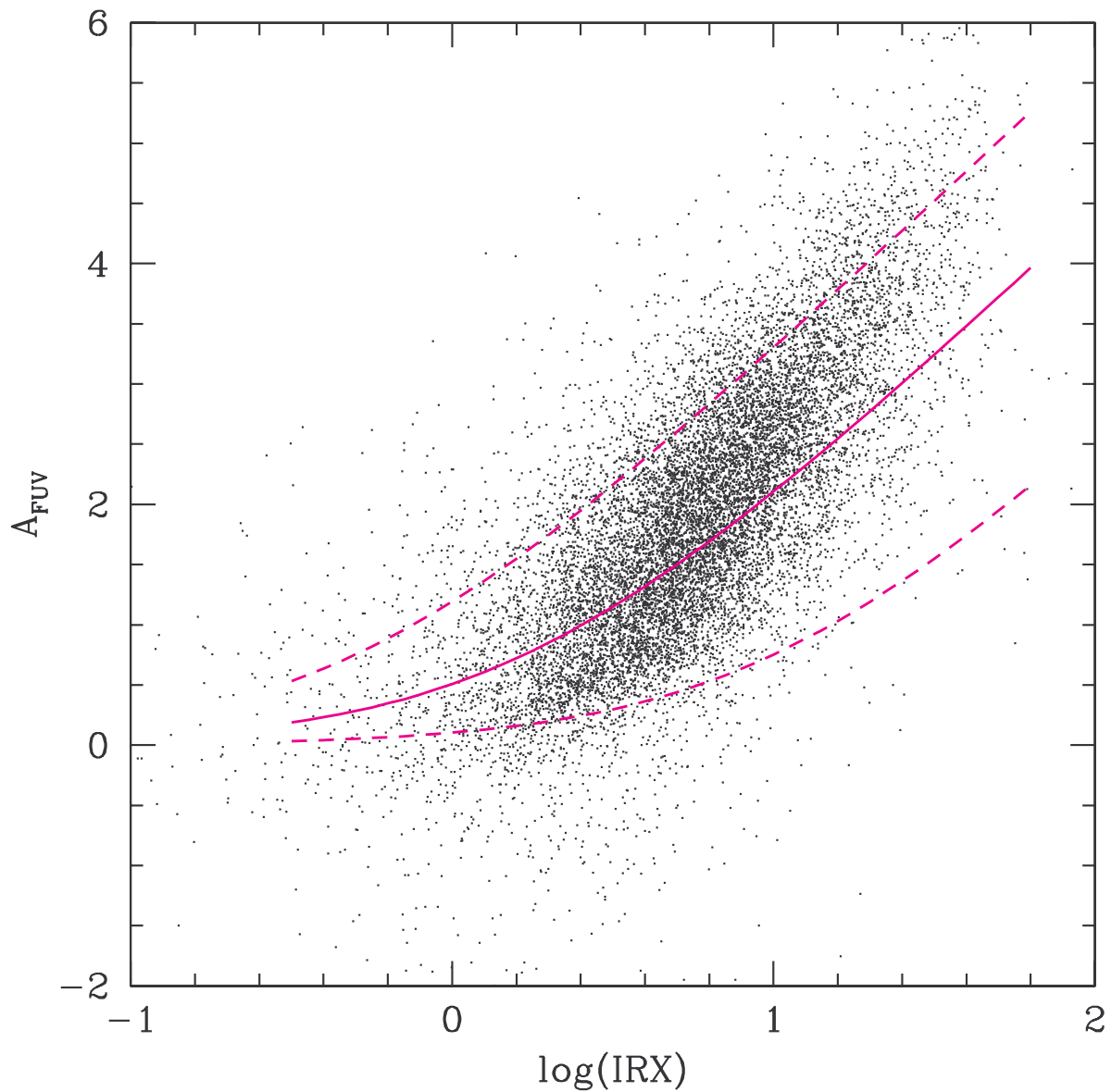}
\caption{The FUV attenuation (Eq. \ref{attenuation}) as a function of IRX derived from D$_n$(4000) and $^{0.1}(n-r)$
following \citet{J07} for the blue population. The solid line is the $A_{FUV}$/IRX relation of \citet{Meurer99} (Eq. \ref{meurer})
assuming $\mu=0.6$.  The lower and upper dotted lines are for $\mu=0.1$ and 2 respectively. 
\label{fig:irx_Afuv}}
\end{figure}

As noted in the previous section, the relation between IRX and $A_{FUV}$ may be responsible for
the discrepancy between  $A_{FUV}$ and $A_{FUV}^{J07}$ (Eq. \ref{ben}).
Figure \ref{fig:irx_Afuv} shows the relation between $A_{FUV}$ (Eq. \ref{attenuation}) and 
IRX derived from D$_n$(4000) and $^{0.1}(n-r)$ (Eq. \ref{meurer} and \ref{ben}). The solid line 
is the relation used by J07 to relate the two quantities (Eq. \ref{meurer} with $\mu=0.6$).
While it is appropriate for the average galaxy, it becomes 
discrepant at the blue and red ends (or for the least and most massive/star-forming galaxies).
There is also a large range of attenuations for a given IRX. 
A color (or SFR) dependent $\mu$ parameter  accounting for the fraction of IR 
flux due to new born stars (as opposed to preexisting older stars) would remedy some of the discrepancy.
A blue galaxy would have little dust attenuation, little FIR emission for a given UV flux (a small IRX) 
and its little FIR emission would seem to have little to do with the new born stars (small $\mu$). 
At the other end, a red (star-forming) galaxy would be a dusty galaxy with a high SFR and 
its large FIR emission (large IRX) would be entirely due to the heating of dust by its
new stars (large $\mu$). A $\mu$ parameter as large as 2 (meaning that the obscured UV emission  
would have to be twice the observed FIR emission) is actually necessary to 
reach the upper enveloppe of the IRX/$A_{FUV}$ distribution. This might be accounted for by 
the uncertainty in the IR flux estimate, or by a large fraction of obscured UV photons being
reprocessed at wavelengths other than IR. J07 modeled the contribution of new born stars
to IRX as a function of galaxy color and arrived at the opposite conclusion: red galaxies
have a higher contribution of older stars to their IR emission than blue galaxies and should
therefore required a lower $\mu$, 
making the trend between the corrected UV luminosity and $SFR_e$ even more pronounced.
The validity of $SFR_e$ at low and high mass may of course be questionned but it seems that
the estimate of the IR flux and the interpretation of IRX in terms of FUV attenuation 
currently involve more uncertainties than the interpretation of the optical data. 

\subsection{Limitations: the oldest and youngest galaxies}

The color corrections do not apply to red sequence galaxies in the local sample. Dust attenuation
estimates based on correlations between IRX and colors do not apply well to early-type
galaxies either, both the UV and IR SEDs of such galaxies being much less directly related to
the emission of young stars than those of late-type galaxies (J07). 
This isn't a drastic limitation to the various methods since red sequence galaxies contribute little to the overall SFR
in the local Universe, and even less as redshift increases. But we would like to know whether the above correlations 
between attenuation and UV--optical color apply to star-forming galaxies at higher redshift, where rest-frame UV fluxes 
are generally corrected using the IRX/$\beta$ correlation of local starburst galaxies \citep{Meurer99}. 

Although no spectroscopic data exist at high redshift that allow the same emission line fitting technique as the SDSS spectra,
we can use the unique sample of \cite{Erb06_Ha} who were able to acquire H$\alpha$ flux measurements for 114 UV selected 
galaxies at $z\sim 2$, for which optical and NIR photometry is also available. 
They defined $SFR_{FUV}$ as in Eq. \ref{sfruv} and $SFR_{{\rm H}\alpha}=L_{{\rm H}\alpha}/\eta_{{\rm H}\alpha}$, 
using the $\eta_{FUV}$ and $\eta_{{\rm H}\alpha}$ values of \cite{Ken98} converted to a \cite{Chabrier03} IMF
\footnote{To do so, Erb et al. multiplied the Kennicutt factors (computed for a Salpeter IMF) by 1.8, however 
a conversion factor of 1.58 between the Salpeter and Chabrier IMFs is more appropriate (S. Salim, private 
communication). We corrected their SFRs accordingly and multiplied them by 0.94 to account for the small 
difference between the Chabrier and Kroupa IMFs.}.
A factor of two aperture correction was also applied to the H$\alpha$ luminosities. 
Dust corrections were derived from the best-fit values of $E(B-V)$ obtained from fitting 
SED models to the multiband photometry and using the extinction law of \cite{Calzetti00} 
(see \cite{Erb06_Ha} for details). The authors assumed that the color 
excess of the nebular emission lines was equal to that of the UV continuum, rather than 2.5 times 
larger as proposed by \cite{Calzetti00}, as it yielded the best agreement between the UV and 
H$\alpha$ SFRs after dust correction. The corrected $SFR_{{\rm H}\alpha}$ are at most
3 times the uncorrected values and less than twice for most of the sample. 

As high redshift galaxies have lower metallicities than local galaxies on average, a higher value 
of $\eta_{{\rm H}\alpha}$ might be justified for this sample.
However as both the range of masses and the range of metallicities span by the $z=2$ galaxies 
remain within those of the SDSS sample despite evolution in the mass/metallicity
relation \citep{Erb06_metals}, $\eta_{{\rm H}\alpha}$ is not expected to be larger than the value 
predicted for the least massive/most metal poor galaxies in the 
local sample, i.e. a factor of 1.5 higher than the Kennicutt value used by Erb et al. 
(B04, their Fig.~7). We assume the uncorrected values of $SFR_{{\rm H}\alpha}$ 
divided by 1.5 to be lower limits to the SFR and use the dust corrected values as upper limits.
We computed the absolute magnitudes of the galaxies in the GALEX and SDSS bands from 
their optical and NIR photometry using {\tt kcorrect v4\_1} \citep{Blanton07_kcorrect}.

\begin{figure}
%\plotone{luv_sfr_col_sb.ps}
\plotone{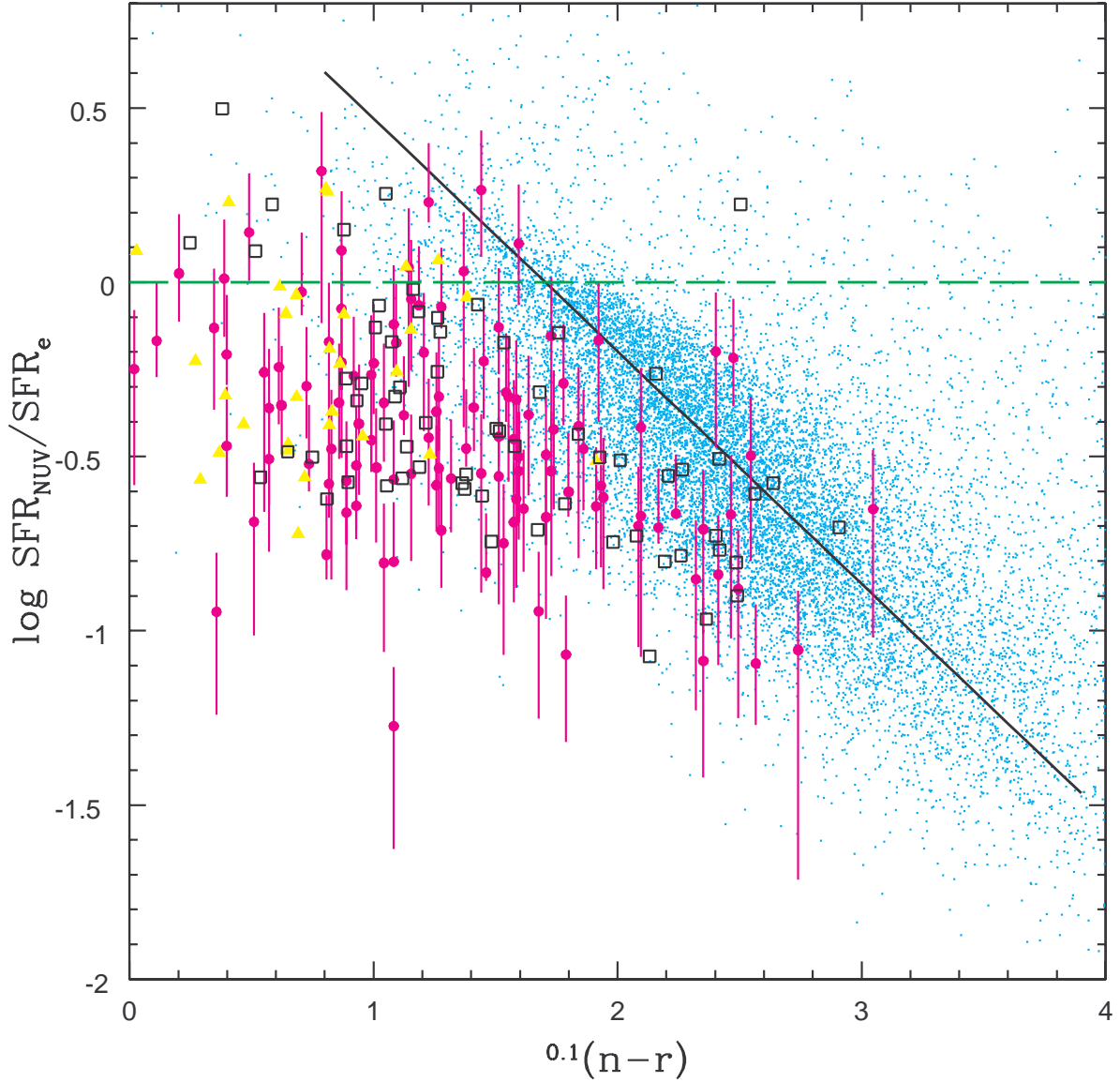}
\caption{The $SFR_{NUV}$ to $SFR_e$ ratio as a function of $^{0.1}(n-r)$ for the local blue population (blue dots), 
the sample of local  compact  and ``supercompact"  UV Luminous Galaxies (UVLGs) of Hoopes et al. (2006) (open 
squares and filled yellow triangles respectively) and the $z\sim2$ Lyman break galaxy (LBG) sample of \citet{Erb06_Ha} 
assuming $SFR_{\rm H\alpha}=SFR_e$ (pink filled circles; see text for details). The straight line is the correlation 
fitted to the local blue population (Eq. \ref{Anuv_nr}). 
\label{fig:luv_sfr_col_sb}}
\end{figure}

Figure \ref{fig:luv_sfr_col_sb} shows $SFR_{NUV}/SFR_e$ as a function of $^{0.1}(n-r)$  
assuming $SFR_{{\rm H}\alpha}=SFR_e$ for the high redshift sample and $\eta_{NUV}=\overline \eta_{FUV}$
for all galaxies as before. The pink circles are the uncorrected values of $SFR_{{\rm H}\alpha}$. 
The upper and lower limits are defined as above. The local blue population is plotted in blue with the 
correlation derived in the previous section (Eq.~\ref{Anuv_nr}). 
It is clear that the majority of LBGs which have extremely blue colors do not follow the same correlation 
as the local galaxies, but the reddest ones ($^{0.1}(n-r)>1.5$) may still be consistent with the local 
correlation or show a similar trend with color with only a small blue shift. 
The bluest LBGs cluster blueward of the local correlation in a region of low UV attenuation.

Also overplotted is a sample of 97 compact UV Luminous Galaxies (UVLGs) drawn 
from the present sample and from a cross-match between the SDSS and the larger, shallower 
GALEX All Sky Imaging Survey \citep{Hoopes06}. $SFR_e$ from B04 are available for all of them. 
UVLGs \citep{Heckman05,Hoopes06} are locally rare galaxies defined as having FUV luminosities typical 
of LBGs: 
$L_{FUV}>2\times 10^{10} L_\odot$, corresponding to $\sim 0.3 L_{\star}$ at $z\sim 3$ \citep{Steidel99}
but to $\sim 5 L_{\star}$ at $z\sim 0$ \citep{Wyder05}. While low surface brightness UVLGs are simply 
extra large versions of normal spiral galaxies, high surface brightness UVLGs 
with $I_{FUV}>10^8 L_\odot {\rm kpc}^{-2}$ were found to consist primarily of compact starburst systems. 
Among these, the ``supercompact'' UVLGs with $I_{FUV}>10^9 L_\odot {\rm kpc}^{-2}$ bear a remarkable 
resemblance to high redshift LBGs for a wide range of physical properties (mass, SFR, metallicity). 
They are thought to be their closest analogs in the Local Universe \citep{Hoopes06}. 
Compact and supercompact UVLGs are represented in Fig.~\ref{fig:luv_sfr_col_sb}
with open squares and filled triangles respectively. Both categories occupy the same region
of the plot as the high redshift sample.
The supercompact UVLGs are unusually blue ($^{0.1}(n-r)<1.5$) among the local galaxies and like 
the bluest LBGs, lie the furthest away from the bulk of the local population. 
The extreme blue colors are an indication of strong recent star-formation (as well as low attenuation).
As noted by \cite{Ken98}, the calibration $\eta_{UV}$ in Eq. \ref{sfruv} might be significantly 
higher for strong starburst galaxies such as these. A higher $\eta_{UV}$ 
would lower their  $SFR_{UV}$ to $SFR_e$ ratio proportionally and further separate them
from the main population. Therefore very blue galaxies with very recent star-formation, both locally 
and at high redshift, form a distinct cluster of their own blueward of the attenuation/color relation 
of the blue sequence. This locus adds to the blue shift of the attenuation/color relation with 
D$_n$(4000) seen in Fig.~\ref{fig:luv_sfr_col} (left panel) as those galaxies would have
D$_n$(4000) indices lower than our lowest bin. The reddest of the compact UVLGs and of the LBGs
cover a wider range of attenuations which seem to correlate with colors in the same way as local 
galaxies only shifted to bluer colors. 

\begin{figure}
%\plotone{col_tsfr.ps}
\plotone{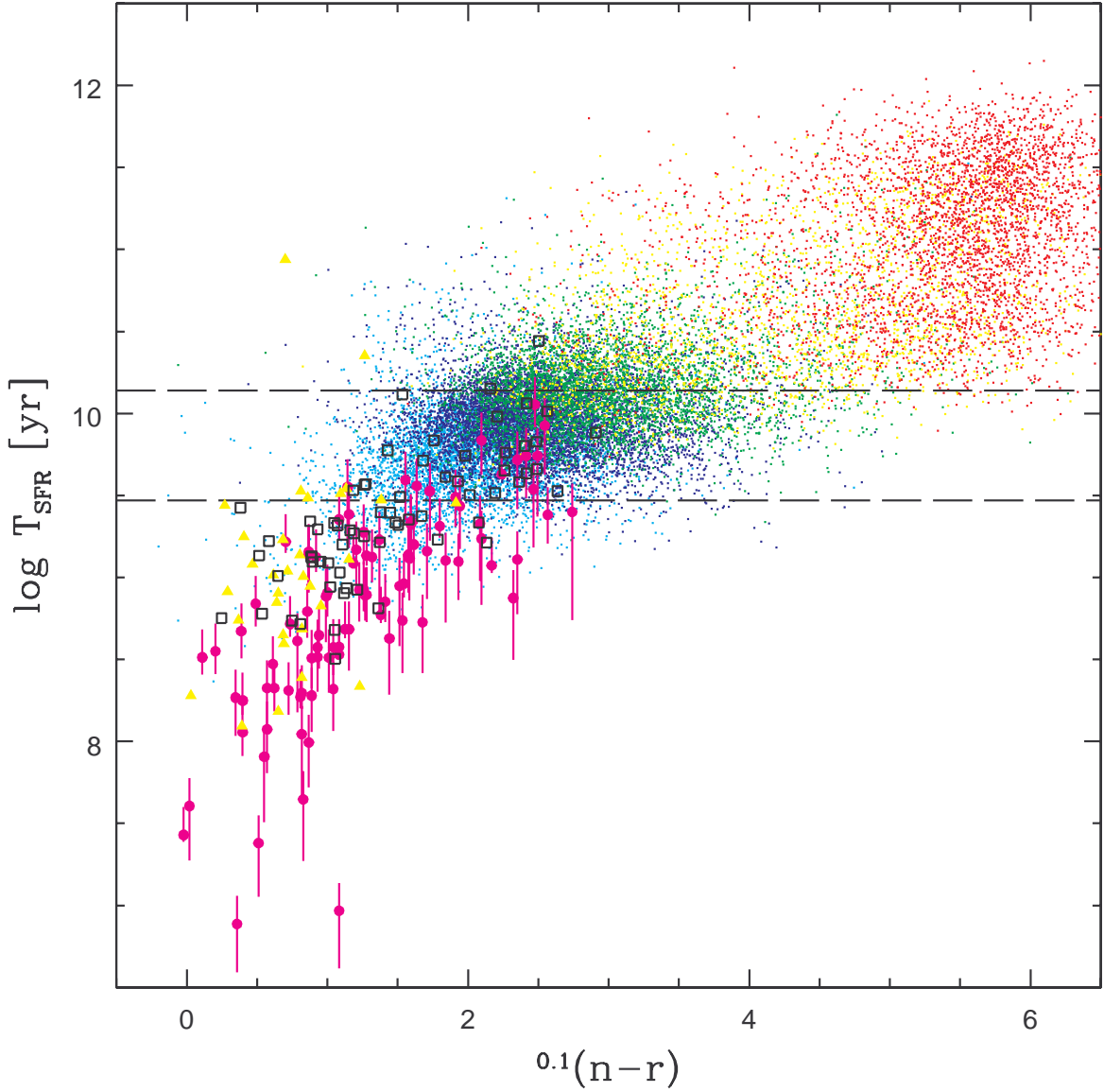}
\caption{The SFR timescale $T_{SFR}=M_\star/SFR_e$ as a function of $^{0.1}(n-r)$.
Local galaxies are shown by dots color coded in bins of D$_n$(4000) as in Fig.~\ref{fig:luv_sfr_col}. 
UVLGs and LBGs are represented as in Fig.~\ref{fig:luv_sfr_col_sb}. The two horizontal dashed lines mark the age 
of the Universe at $z=2$ and $z=0$ (3 and 13.8 Gyrs respectively). This plot is similar to Fig. \ref{fig:cold4000} 
with the addition of an ultrablue sequence at $^{0.1}(n-r)<1.5$ and $T_{SFR}<3$ Gyrs consisting of young compact 
starburst galaxies for which the local UV attenuation/UV--optical color relations derived in this paper do not apply.
\label{fig:col_tsfr}}
\end{figure}

D$_n$(4000) estimates are not available for the high redshift sample and only available for a fraction
of the UVLGs but we can use the inverse of the specific SFR -- $T_{SFR}=M_\star/SFR$ -- as a common SFR 
time-scale for the local and high redshift samples. Stellar masses were estimated for most of the 
$z\sim2$ sample \citep{Erb06_masses}. Masses for the GALEX/SDSS sample are from \cite{Kauffmann03_masses}.
A third of the UVLGs have mass estimates from this catalog; for the remaining 2/3 we use the values
derived by \cite{Hoopes06} via SED fitting following \cite{Salim05}. The agreement between the mass 
estimates of \cite{Kauffmann03_masses} and \cite{Salim05} for the GALEX/SDDS sample is good with 
a rms of 0.12.
As above, we use $SFR=SFR_e$ for the local sample including the UVLGs and 
$SFR_{{\rm H}\alpha}$ for the $z\sim2$ galaxies.
Figure \ref{fig:col_tsfr} shows $T_{SFR}$ as a function of $^{0.1}(n-r)$.
The SDSS galaxies are color coded in bins of D$_n$(4000) as shown in the inset of
Fig.~\ref{fig:luv_sfr_col} (this highlights the relation between D$_n$(4000) and $T_{SFR}$).
The LBGs are represented by filled circles, the compact UVLGs by open squares and the
supercompact UVGLs by filled triangles.
The two horizontal lines correspond to the age of the Universe at $z=0$ and 2
(13.8 Gyrs and 3 respectively). Roughly,
galaxies with $T_{SFR}$ larger than the age of the Universe at their redshift
(e.g. nearly all local galaxies with D$_n$(4000)$>1.6$ and a few LBGs) 
have had larger SFR in the past. Inversely, galaxies with $T_{SFR}$ shorter 
than the age of the Universe at their redshift (most LBGs and nearly all the compact UVLGs) 
must be forming stars more intensely than in the past.

Figure \ref{fig:col_tsfr} is an analog of Fig. \ref{fig:cold4000} with the addition of an `ultrablue' sequence 
at $^{0.1}(n-r)<1.5$ and $T_{SFR}<3$ Gyrs consisting of young compact starburst galaxies.
The $A_{UV}$/color correlations derived in the previous section hold for 
galaxies with rather uneventful SFHs (3 Gyrs $\la T_{SFR} \la 15$ Gyrs). They are a majority today 
but may not be when the Universe was only $\sim$ 3 Gyrs old, although many LBGs 
look very much like local blue sequence galaxies. Furthermore a dominant fraction of the stellar mass at $z>2$ 
is found in redder galaxies which are largely absent from UV surveys \citep{Rudnick06,vanDokkum06,Marchesini07}. 
\citet{Kriek06_NIRspec} showed that almost half of their sample of NIR selected galaxies at $z=2.0-2.7$
have low SFRs  and $T_{SFR} > 10$ Gyr (from their Fig. 2). These galaxies are redder than LBGs
and would lie in the same part of the plot as the local population. 

Although we can't conclude on the use of the local $A_{UV}$/color correlation at high redshift, 
we may expect it to hold for most galaxies to at least  intermediate redshifts or to be slightly 
shifted to the left as only `mild evolution' of the blue sequence was reported between $z=0$ and 1 
(no change in the number density and colors only $\sim 0.3$ mag bluer; Blanton 2006 \nocite{Blanton06}).
The luminosity density from UVLGs was found to undergo dramatic evolution between $z=0$ and 1, reaching $>25\%$ of 
the total FUV luminosity density at $z=1$ \citep{Schimi05}, but this includes
all UVLGs, i.e. mostly very large but otherwise ordinary spiral galaxies. Compact and supercompact 
UVLGs are very rare in the local Universe. Although their evolution with redshift is yet unknown,
they are unlikely to dominate the galaxy population at $z=1$. 

\section{Conclusions}

Using a large sample of galaxies from the Sloan Digital Sky Survey spectroscopic catalog
with measured SFRs and UV photometry from the GALEX Medium Imaging Survey, we derived empirical 
linear correlations between the UV attenuation measured by the SFR to observed UV luminosity ratio, 
and the UV$-$optical colors of blue sequence galaxies ($^{0.1}(n-r)<4$). The SFRs were derived from 
a detailed modelling of the emission lines in the optical spectra \citep{B04} and were 
considered best estimates. 
The attenuation/color relation provides a simple prescription to correct UV measurements for dust 
attenuation in the absence of SDSS quality data. We found or confirmed that other UV attenuation 
estimates (Seibert et al. 2005, Salim et al. 2007, Johnson et al. 2007) tend to over (under)
correct the UV luminosity of galaxies with the lowest (highest) emission line SFRs or mass.
Using a sample of LBGs \citep{Erb06_Ha} at $z\sim 2$ with measured H$\alpha$ emission as well as 
a sample of local compact UV luminous galaxies with LBG like properties \citep{Hoopes06}, we found 
that extremely blue galaxies at both low and high redshift escaped the attenuation/color relation 
of the blue sequence to form a low attenuation sequence of their own. As such galaxies
are very rare locally and the blue sequence does not evolve much from $z=0$ to 1 \citep{Blanton06,Willmer06},
we expect our attenuation correction to remain adequate for the majority of galaxies to at least
intermediate redshifts.

\acknowledgments   
The authors thank their anonymous referee whose many comments greatly improved this paper.
GALEX is a NASA Small Explorer, launched in April 2003. We gratefully acknowledge 
NASA's support for its construction, operation, and science analysis, as well as the
cooperation of the French Centre National d'Etudes Spatiales and of the Korean Ministry 
of Science and Technology. 

%\bibliographystyle{apj}
%\bibliography{galexref}

\clearpage

\begin{deluxetable}{cccccc}
\tablecolumns{6}
\tablewidth{0pc}
\tablecaption{Correlation parameters between the corrected FUV SFR ($SFR_{FUV,model}$) and $SFR_e$ for the various attenuation 
models described in the text as listed in the first column. The next columns list the slope $a$ of the linear correlation fitted to the $SFR_{FUV,model}$
 vs  $SFR_e$ relation, the variance of the slope, the correlation coefficient, the rms and the average of the  $SFR_{FUV,model}/ SFR_e$ ratios.
\label{slopes}}
\tablehead{
\colhead{Model} & \colhead{slope $a$} & \colhead{Var($a$)} & \colhead{$r$} & \colhead{rms} & \colhead{ $<SFR_{FUV.model}/SFR_e> $} }
\startdata
%      model           a       Var_a           r         rms        mean      
        Se05 &     0.9498 &  5.181e-05 &     0.7392 &     0.4209  &   0.0012   \\
        Sa07 &     0.8314 &   2.23e-05 &     0.7866 &     0.3328  &   0.0421   \\
         J07 &     0.7171 &  1.242e-05 &     0.8658 &     0.2278  &  -0.0847   \\
       Eq. 9 &     0.9799 &   5.14e-05 &     0.7487 &     0.4266  &   0.0061   \\
       Eq. 8 &     0.9258 &  1.915e-05 &     0.8705 &     0.2893  &   0.0316   

\enddata
\end{deluxetable}

\end{document}